\documentclass{pasa}%
\usepackage{graphicx}
\usepackage[usenames,dvipsnames]{color}
\title[An ultra-wide bandwidth (704 to 4032\,MHz) receiver for the Parkes radio telescope]{An ultra-wide bandwidth (704 to 4032\,MHz) receiver for the Parkes radio telescope}
\author[Hobbs, G. et al.]{
G.~Hobbs$^1$\thanks{george.hobbs@csiro.au}, 
R.\,N.~Manchester$^1$, 
A.~Dunning$^1$,
A.~Jameson$^2$,
P.~Roberts$^1$,
D.~George$^1$, 
J.\,A.~Green$^3$, 
J.~Tuthill$^1$,
L.~Toomey$^1$,
J.\,F.~Kaczmarek$^{1,4}$, 
S.~Mader$^4$,  
M.~Marquarding$^1$,  
A.~Ahmed$^1$,
S.~W. Amy$^1$,
M.~Bailes$^{2,6}$,
R.~Beresford$^1$, 
N.\,D.\,R.~Bhat$^5$, 
D.\,C.-J.~Bock$^1$, 
M.~Bourne$^1$,
M.~Bowen$^1$,
M.~Brothers$^1$,
A.\,D.~Cameron$^1$,
E.~Carretti$^{7}$, 
N.~Carter$^1$, 
S.~Castillo$^1$, 
R.~Chekkala$^1$, 
W.~Cheng$^1$,
Y.~Chung$^1$, 
D.\,A.~Craig$^1$, 
S.~Dai$^1$, 
J.\,R.~Dawson$^8$, 
J.~Dempsey$^{9,10}$, 
P.~Doherty$^1$, 
B.~Dong$^{11}$,
P.\,G.~Edwards$^1$, 
T.~Ergesh$^{12}$,
X.\,Y.~Gao$^{11}$,
J.\,L.~Han$^{11}$, 
D.\,B.~Hayman$^1$, 
B.\,T.~Indermuehle$^1$, 
K.~Jeganathan$^1$,
S.~Johnston$^1$, 
H.~Kanoniuk$^1$,
M.~Kesteven$^1$, 
M.~Kramer$^{13}$,
M.~Leach$^1$,
V.\,J.~Mcintyre$^1$,
V.\,A.~Moss$^{1,14,16}$, 
S.~Os{\l}owski$^2$, 
C.\,J.~Phillips$^1$,  
N.\,C.~Pope$^1$,
B.~Preisig$^4$, 
D.\,C.~Price$^2$, 
K.~Reeves$^4$,
L.~Reilly$^1$,
J.\,E.~Reynolds$^1$,
T.~Robishaw$^{15}$,
P.~Roush$^1$, 
T.~Ruckley$^4$,
E.\,M.~Sadler$^{1,16}$,
J. Sarkissian$^4$,
S.~Severs$^1$,
R.\,M.~Shannon$^{2,6}$,
K.\,W.~Smart$^1$,
M.~Smith$^4$, 
S.\,L.~Smith$^1$,
C.~Sobey$^{3}$,  
L.~Staveley-Smith$^{17}$,
A.\,K.~Tzioumis$^1$, 
W.~van~Straten$^{18}$, 
N.~Wang$^{12}$,
L.~Wen$^{19}$,
M.~T.~Whiting$^1$ 
\\~\\
\affil{$^1$ CSIRO Astronomy \& Space Science, Australia Telescope National Facility, P.O. Box 76, Epping, NSW 1710, Australia} 
\affil{$^2$ Centre for Astrophysics and Supercomputing, Swinburne University of Technology, P.O. Box 218, Hawthorn, Victoria 3122, Australia}
\affil{$^{3}$ CSIRO Astronomy and Space Science, P.O. Box 1130, Bentley, WA 6102, Australia}
\affil{$^4$ CSIRO Astronomy and Space Science, Parkes Observatory, P.O. Box 276, Parkes NSW 2870, Australia}
\affil{$^5$ International Centre for Radio Astronomy Research, Curtin University, Bentley, WA 6102, Australia}
\affil{$^6$ ARC Centre of Excellence for Gravitational Wave Discovery (OzGrav)}
\affil{$^7$ INAF - Istituto di Radioastronomia, Via Gobetti 101, 40129 Bologna, Italy}
\affil{$^8$ Department of Physics and Astronomy and MQ Research Centre in Astronomy, Astrophysics and Astrophotonics, Macquarie University, NSW 2109, Australia}
\affil{$^9$ CSIRO Information Management and Technology, GPO Box 1700 Canberra, ACT 2601, Australia}
\affil{$^{10}$ Research School of Astronomy \& Astrophysics, Australian National University, Canberra, ACT 2611, Australia}
\affil{$^{11}$ National Astronomical Observatories, Chinese Academy of Sciences, A20 Datun Road, Chaoyang District, Beijing, 100101, China}
\affil{$^{12}$ Xinjiang Astronomical Observatory, Chinese Academy of Sciences, 150 Science 1-Street, Urumqi, Xinjiang 830011}
\affil{$^{13}$ Max-Planck-Institut f\"{u}r Radioastronomie, Auf dem H\"{u}gel 69, D-53121 Bonn, Germany}
\affil{$^{14}$ ASTRON, The Netherlands Institute for Radio Astronomy, Postbus 2, NL-7900 AA Dwingeloo, the Netherlands}
\affil{$^{15}$ National Research Council Canada, Herzberg Astronomy and Astrophysics Programs, Dominion Radio Astrophysical Observatory, PO Box 248, Penticton, BC V2A 6J9, Canada}
\affil{$^{16}$ School of Physics, Sydney Institute for Astronomy, The University of Sydney, Sydney NSW 2006, Australia}
\affil{$^{17}$ International Centre for Radio Astronomy Research, University of Western Australia, Crawley, WA 6009, Australia}
\affil{$^{18}$ Institute for Radio Astronomy \& Space Research, Auckland University of Technology, Private Bag 92006, Auckland 1142, New Zealand}
\affil{$^{19}$ Department of Physics, University of Western Australia, Crawley, WA 6009, Australia}
}
\jid{PASA}
\doi{10.1017/pas.\the\year.xxx}
\jyear{\the\year}
\usepackage{aas_macros}
\usepackage[colorlinks=true,citecolor=blue,linkcolor=blue]{hyperref} 
\begin{document}
\begin{frontmatter}
\maketitle
\begin{abstract}
We describe an ultra-wide-bandwidth, low-frequency receiver (`UWL') recently installed on the Parkes radio telescope. The receiver system provides continuous frequency coverage from 704 to 4032\,MHz. For much of the band ($\sim$60\%) the system temperature is approximately 22\,K and the receiver system remains in a linear regime even in the presence of strong mobile phone transmissions.   We discuss the scientific and technical aspects of the new receiver including its astronomical objectives, as well as the feed, receiver, digitiser and signal-processor design. We describe the pipeline routines that form the archive-ready data products and how those data files can be accessed from the archives. The system performance is quantified including the system noise and linearity, beam shape, antenna efficiency, polarisation calibration and timing stability. 
\end{abstract}
\end{frontmatter}

\section{INTRODUCTION}
\label{sec:intro}

\begin{figure*}
\begin{center}
\includegraphics[width=13cm]{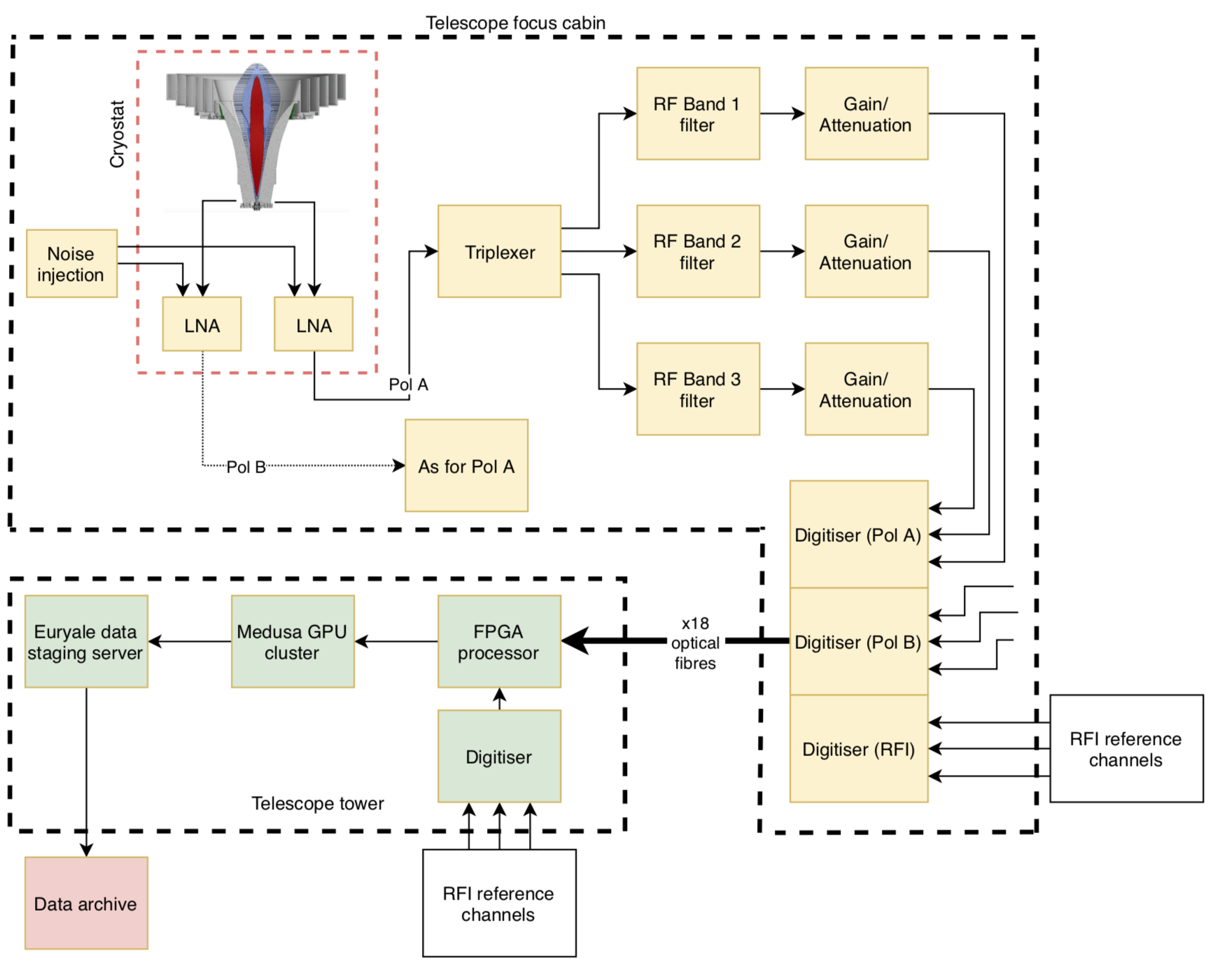}
\caption{Block diagram of the primary components of the UWL system. Only one polarisation channel is shown for the RF amplifier chain; the other is identical.  Note that this figure does not include the monitoring, control nor timing/synchronisation system.}
\label{fg:overview}
\end{center}
\end{figure*}

The Parkes 64-m diameter telescope remains a cutting-edge instrument because of the numerous upgrades that have occurred since it was built in 1961 (see, for example, \citealt{edw12}).  Astronomical requirements have continued to push for receivers that have wider fields of view and/or wider observing bandwidths. Multibeam receivers, such as the 13-beam 20\,cm receiver \citep{swb+96} and the 7-beam ``methanol'' receiver \citep{green2009}, have successfully carried out large-scale survey observations. More recently, an Australian Square Kilometer Array Pathfinder (ASKAP) phased array feed has been productively trialled at Parkes \citep{dch+17,rey17}.

A dual-band receiver has been the primary receiver system for the Parkes Pulsar Timing Array project (PPTA; \citealt{mhb+13}), which carries out high precision timing observations of pulsars in the hunt for ultra-low-frequency gravitational waves.  
That receiver, installed in 2003, initially covered the 653--717\,MHz (50\,cm) and 2.6--3.6\,GHz (10\,cm) bands and so was dubbed the `10/50' receiver. As a result of increasing radio-frequency interference (RFI) in the 50\,cm band, the receiver was re-tuned to cover the 700--764\,MHz (40\,cm) band in 2009, and so was renamed the `10/40' receiver.  The dual-band nature of the receiver enables accurate determinations of pulsar dispersion measure variations (essential for the detection of low-frequency gravitational waves, e.g., \citealt{kcs+13}), but misses the key 20\,cm observing band and leads to incomplete frequency coverage of the pulsar observations (see for example \citealt{dhm+15}). This gap was partially filled by observations using the central beam of the 13-beam 20\,cm receiver.  The dual-band receiver has also been used for continuum studies \citep{car10} 
although low frequency spectroscopic observations have primarily used the 20\,cm multi-beam (1.2--1.5\,GHz) and H-OH (1.2--1.8\,GHz) receivers. 

With its frequency coverage of 704 to 4032\,MHz, the new ultra-wide-bandwidth low-frequency receiver (UWL) described here covers, and extends, the combined frequency ranges of these three existing receiver packages. It is the first in a suite of new receivers planned for Parkes that will provide continuous frequency coverage from $\sim$700\,MHz to $\sim$24\,GHz (hence the ``low-frequency'' qualifier in the name). Although other radio observatories have already installed ultra-wide-bandwidth receivers\footnote{Very large bandwidth receivers exist in relatively high-frequency observing bands and receiving systems with large fractional bandwidth exist at low radio frequencies. Here we discuss cm-wavelength receivers with large fractional bandwidths.}, none have the large fractional bandwidth and low system temperature of the UWL.  A receiver covering from 600 to 3000\,MHz was installed on the Effelsberg telescope\footnote{\url{https://www3.mpifr-bonn.mpg.de/staff/pfreire/BEACON.html}} and demonstrated the requirement for the entire system to remain in a linear operating regime in the presence of strong, out-of-band RFI. The first discoveries \citep{qian19} made with the Five Hundred Metre Aperture Spherical Telescope (FAST) were obtained with a receiver of similar design covering from 300\,MHz to 1.6\,GHz.  However, that receiver system was not cryogenically cooled and provided optimal performance only at the low end of the band.

The primary scientific goals for the UWL were originally envisioned to be tests of theories of relativistic gravitation including the search for gravitational waves (a review is provided in \citealt{hd17}), probing neutron star interiors (such as \citealt{dpr+10}) and investigations into the magnetic field structure of our Galaxy  (for example, \citealt{hmvd18} and \citealt{ccs+13}). Furthermore, the receiver enables numerous and diverse science projects, including those relating to high precision pulsar timing, studying the broad-band nature of pulsar profiles and discovering new pulsars and transient sources. It enables simultaneous observations of the low (722.49 and 724.79\,MHz\footnote{Unfortunately the 701.68 and 703.98\,MHz lines lie just below the band edge.}) and high-frequency (3263.79, 3335.47 and 3349.19\,MHz) methylidyne radical (CH) transitions, hydroxyl (OH) transitions (1612.23, 1665.40, 1667.36, 1720.53\,MHz) and a number of recombination lines. The neutral hydrogen (H{\sc i}) line at 1420.41\,MHz is covered at its rest frequency through to a redshift of $\sim$1. Detections of red-shifted H{\sc i} absorption in this previously largely inaccessible band with ASKAP have demonstrated the advantages of broad frequency coverage (e.g., \citealt{all15}). Similarly, the UWL receiver can be used for studies of extra-galactic OH masers \citep[e.g.,][]{2002AJ....124..100D}. Furthermore, the overlapping frequency coverage between the UWL and ASKAP allows Parkes to provide the critical large-angular-scale structure (zero-spacing) data sets to complement the interferometric datasets produced by ASKAP (which has a minimum baseline of 22\,m). The UWL enables observations requiring Very Long Baseline Interferometry (VLBI) in conjunction with a diverse range of national and international telescopes. 

In this paper, we describe the recent upgrade to the Parkes telescope with the UWL and its various components. The receiver and its associated signal-processing systems were installed on the telescope and commissioned during 2018. 

The primary components of the system are shown schematically in Figure~\ref{fg:overview}.   The Parkes telescope is a prime focus system with a focal ratio of 0.41.  The focus cabin includes a ``translator system'' that enables a specific receiver system to be placed on axis.  The UWL receiver system is mounted on a plate that also has space available for future high-frequency receivers (that will extend the frequency coverage to $\sim$24 GHz).  It includes the cryogenically cooled feed and low-noise amplifiers (LNAs), the noise injection system, the radio-frequency (RF) amplifier chain and, for the first time at Parkes, digitisers that sample the entire RF band for each polarisation and stream this data over high-speed serial links, transported on single-mode optical fibres to the telescope tower. 

The signal pre-processor system in the tower receives the digitised data and produces critically-sampled data streams for 26 sub-bands, each with a 128\,MHz bandwidth.  The sub-band data streams are passed to an astronomy signal-processor system based on Graphics Processor Units (GPUs) known as ``Medusa''\footnote{So named for the multiple applications the system can provide, akin to many snakes on the mythological Medusa's head.} that processes each of the sub-bands separately.  The processing can involve forming polarisation products, folding these signals  at the known period of a pulsar, producing data streams suitable for pulsar and transient searching, or producing high frequency resolution data sets for the study of spectral lines and the radio-continuum background.  The output of the astronomy signal processor is transferred to a data-staging server, known as ``Euryale''\footnote{As a sibling to the main system.}, that produces archive-ready data products.  Astronomers can access the resulting data products from various on-line archives.

 The paper is divided into sections describing the receiver system (\S\ref{sec:receiver}), the signal processing system (\S\ref{sec:signalProcessor}), the timing and synchronisation system (\S\ref{sec:timing}), the system performance (\S\ref{sec:systemPerformance}) and the RFI environment (\S\ref{sec:rfi}).  In the Appendices we define the terms used in our paper, describe the content of a publicly downloadable data collection and provide more details on the parameterisation of the receiver system. Finally we briefly present the author contributions to this work.

\begin{figure*}
\includegraphics[width=17cm]{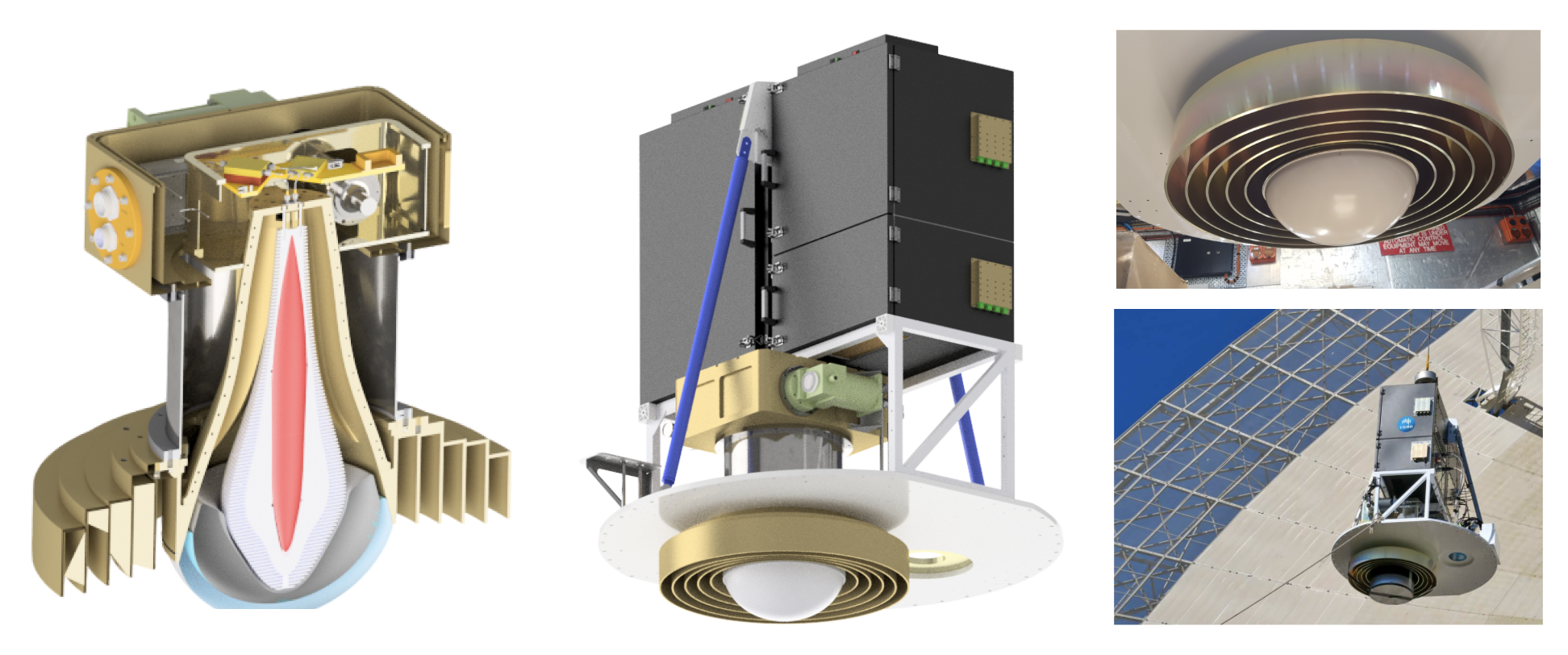}
\caption{Representation of the feed system. (left) the internal structure of the feed, (centre) the feed, platform and radio frequency shielded cabinets and (right) the completed system being installed on the telescope.}
\label{fg:feed}
\end{figure*}


\section{THE RECEIVER SYSTEM}\label{sec:receiver}

The receiver system consists of the feed and LNAs, both of which are cryogenically cooled, and the RF amplifier chain. \cite{dbb+15} described the basic properties of the feed.   The design is shown in the left-hand panel of Figure~\ref{fg:feed} and is based on a quad-ridged horn. The middle and right-hand panels of this Figure show different visualisations of the complete receiver package. This design provides a wide bandwidth with comparable performance to narrower-band systems by making use of a central dielectric spear to improve the beam properties at high frequencies, and a corrugated skirt to improve the beam properties at low frequencies. 

\begin{figure*}
\begin{center}
    \includegraphics[width=9cm,angle=-90]{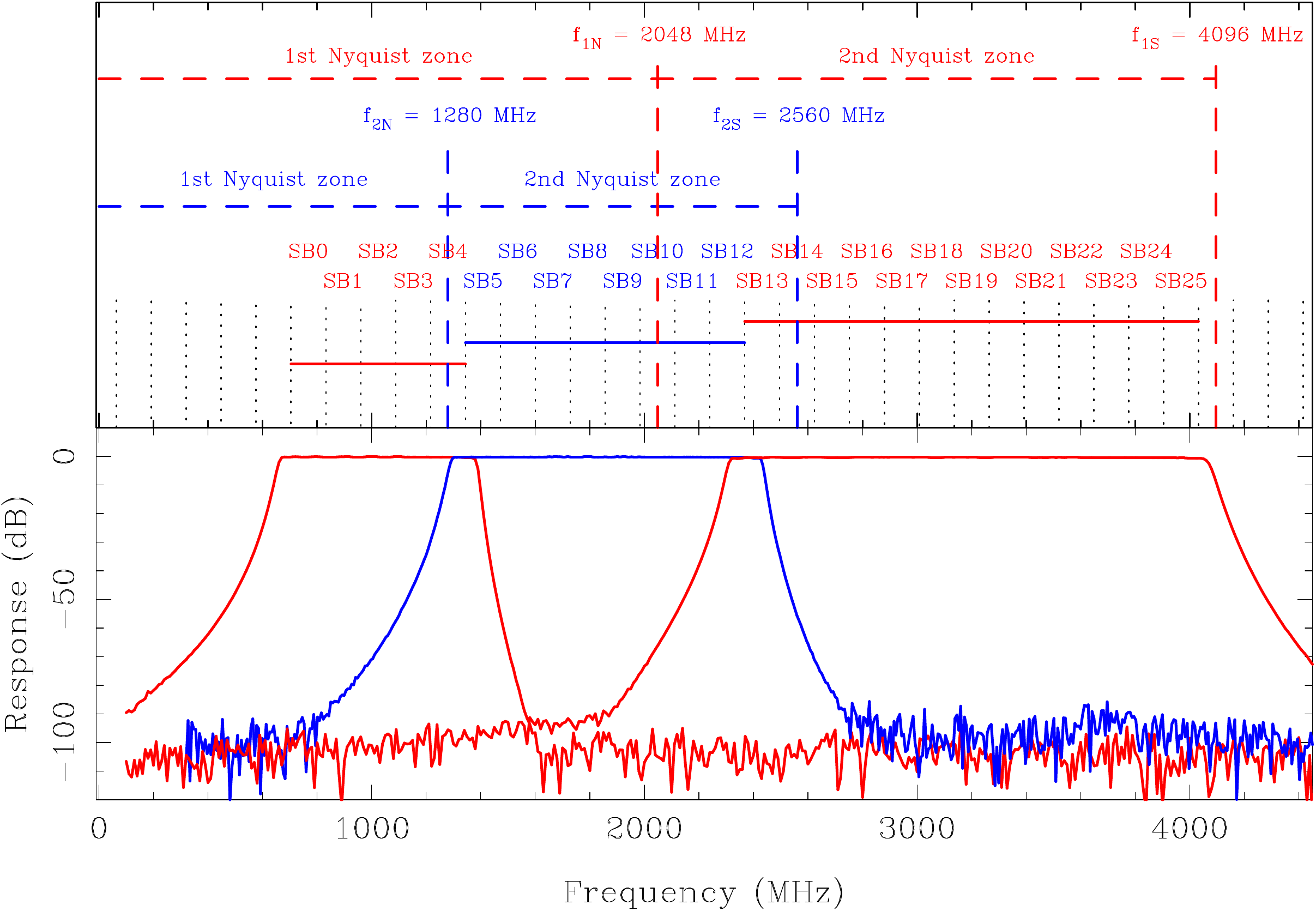}
    \caption{The sampling bands for the UWL. The bottom panel shows the filters determining the three RF bands (RF bands 1, 2 and 3). The digital bands are formed using two sampling frequencies ($f_{1S} = 4096$\,MHz and $f_{2S} = 2560$\,MHz). These sampling frequencies permit the 26 sub-bands (denoted SBx) each 128\,MHz wide. Those shown in red are obtained from the 1st and 2nd Nyquist zones of $f_{1S}$. Those in blue are from the 2nd Nyquist zone of $f_{2S}$. See text for further details.}
    \label{fg:samplebands}
    \end{center}
\end{figure*}

The dielectric spear consists of three layers giving a graded dielectric constant: a central quartz section, a solid polytetrafluoroethylene (PTFE; more commonly known as Teflon$^{\rm TM}$) section and a slotted PTFE outer section.  The feed horn operates over the frequency range $\sim$700\,MHz to $\sim$4.5\,GHz and provides two nominally orthogonal linear polarisations, each with symmetric outputs. The exceptionally wide frequency coverage of the feed required the development of cryogenically cooled LNAs with large dynamic range that cover the entire band. The three-stage LNAs are cooled to 20\,K, the feed horn and the dielectric spear are cooled to 70\,K and the outer feed rings are at ambient temperature (because of thermal losses, there is a temperature gradient across the dielectric spear; see \citealt{sds+19} for more detail).
Following the cryogenically cooled LNAs, the signals from the two polarisations are further amplified by ambient-temperature amplifiers before digitisation. To assist with the mitigation of RFI, the UWL band for each polarisation is split into three sub-octave RF bands using a triplexer (created using a power splitter and a diplexer) and band-limiting filters. These bands are separately amplified and digitised. As shown in the bottom panel of Figure~\ref{fg:samplebands}, the three RF bands (labelled Bands 1 to 3) have a relatively flat response over the corresponding digital bands, which cover from 704--1344\,MHz, 1344--2368\,MHz and 2368--4032\,MHz (see \S\ref{sec:signalProcessor} for more detail on the digital bands). The separate sub-octave RF bands provide protection against inter-modulation products from strong RFI signals affecting other parts of the UWL band. Switchable attenuators allow the choice of signal level for the three RF bands, with RF Band 1 requiring significantly more attenuation than the other two because of the strong RFI in this band.  Default attenuator settings ensure that each band normally remains in a linear regime in the presence of RFI\footnote{Dynamic range measurements show that, in the low band, the 1\,dB compression point of the RF amplifier chain is about 20\,dB above the operating power with standard attenuator settings. This is generally sufficient to avoid intermodulation products from strong RFI signals; see Section~\ref{sec:rfi}.}.

A key aspect of the design was to digitise the signals at RF in the focus cabin and then to transmit digital signals to the telescope tower that contains the signal-processor instrumentation. The alternative of using analogue RF-over-fibre for the long ($\sim 150$~m) path to the tower was rejected because of potential linearity issues related to the strong RFI, particularly for RF Band 1.
Also, analogue transmission over such a long path would be subject to differential delays in the two polarisations with varying ambient conditions, leading to instabilities in derived polarisation parameters. 

 The UWL cannot be rotated (both software and hardware limits are in place to ensure this) and the linear polarisations have been aligned at $\pm$45$^\circ$ to the elevation axis of the antenna. The UWL feed design precludes the injection of noise into the feed horn. The artificial noise signal is therefore coupled  directly to the inputs of the LNAs. Note that the coupling is non-directional; see Section~\ref{sec:systemPerformance}. We also have a capability to radiate noise directly into the feed from a transmitter on the surface of the dish. However, this is currently not used in any standard observing mode.
 It is possible to vary the amplitude of the injected signal; for the results presented here, the amplitude was between 1 and 2\,K across the band. 

\begin{figure*} 
\begin{center}
\includegraphics[angle=-90,width=17cm]{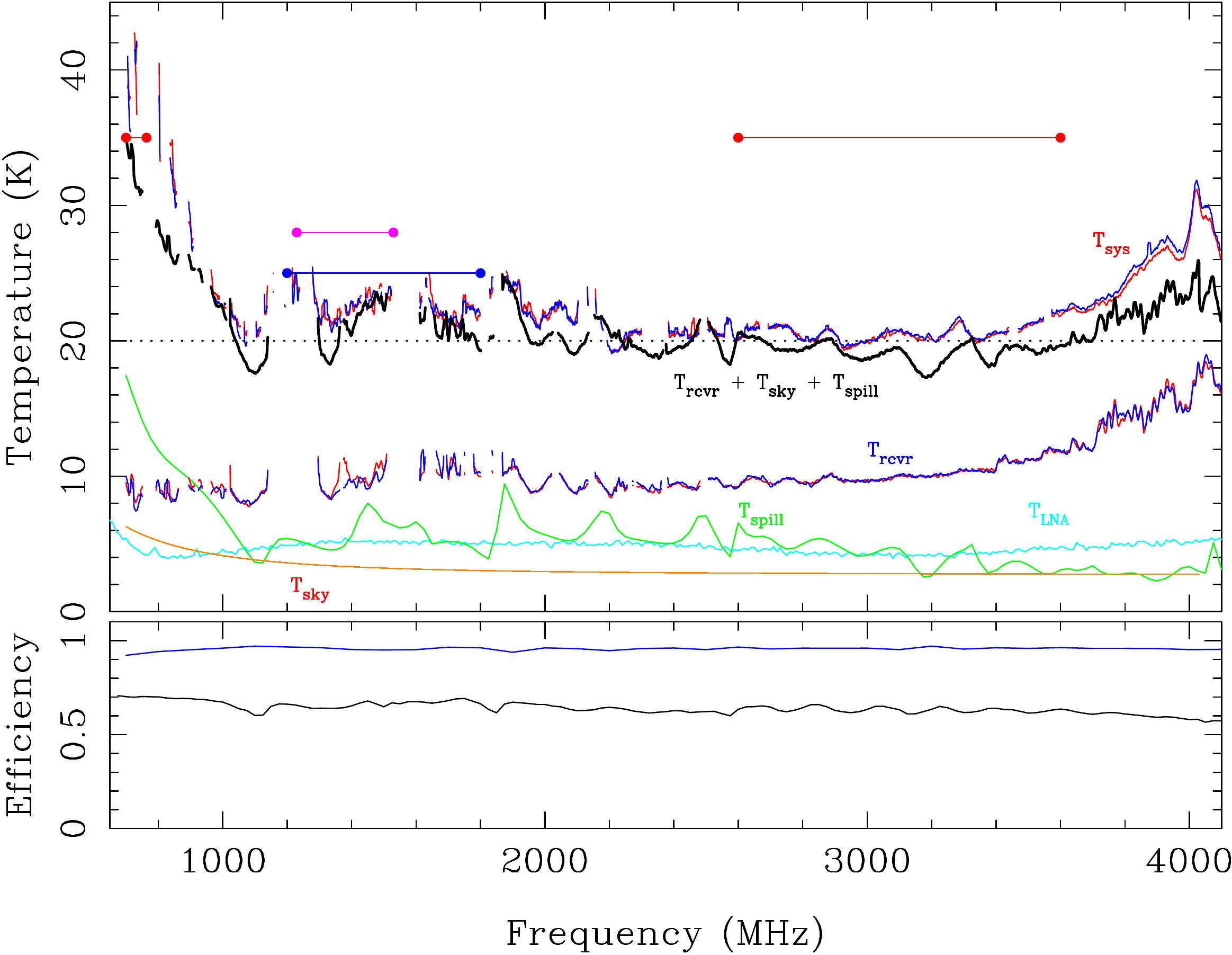}
\caption{The measured system temperature (T$_{\rm sys}$) for the UWL receiver system on the Parkes radio telescope and several of its components are shown in the upper panel. The red and dark blue traces indicate the A and B polarisations respectively for $T_{\rm sys}$ (upper pair) and $T_{\rm rcvr}$ (lower pair). The solid, cyan line ($\sim 5$\,K) is the measured $T_{\rm LNA}$, the green line is the estimated $T_{\rm spill}$ and the orange curve ($< 5$\,K) is the estimated sky background temperature, $T_{\rm Gal} + T_{\rm CMB}$. The upper, black line is the sum of the individual estimated noise components. The horizontal dotted line indicates 20\,K. The horizontal lines delineated by circles give an approximation to the frequency coverage and system temperature for other receivers available at Parkes. The 10/40cm receiver is represented in red, the 20\,cm multibeam receiver in pink and the H-OH receiver in blue. The lower panel provides measures of the aperture efficiency from laboratory-based measurements and simulation (black line). The higher, blue line in this panel provides an estimate of the main beam efficiency.} 
\label{fg:tsys}
\end{center}
\end{figure*}

The noise temperature of the LNAs, $T_{\rm LNA}$, was determined using laboratory measurements in a cryogenic test dewar. The results are shown as the solid, cyan line in the upper panel of Figure~\ref{fg:tsys} and is close to 5\,K across the entire band. The system temperature (note that, as we are not attempting a rigorous analysis here, we ignore efficiencies and beam-pattern weightings) is the sum of several components, which are individually defined below:
\begin{eqnarray}
    T_{\rm sys} &=& T_{\rm rcvr} + T_{\rm atm} + T_{\rm Gal} + T_{\rm CMB} +  \\ \nonumber
    & & T_{\rm spill} + T_{\rm scatt} + T_{\rm RFI}.
\end{eqnarray}\label{eq:tsys}
 The $T_{\rm rcvr}$ measurements (which include $T_{\rm LNA}$), also shown in the top panel of Figure~\ref{fg:tsys}, were made at the Parkes observatory site with the receiver on the ground facing up, using an ambient-temperature absorber as a hot load and the sky as a cold load.  $T_{\rm rcvr}$, which includes contributions from losses in the feed, is  $\sim 9$\,K across most of the band, rising from around 3\,GHz to $\sim 18$\,K at $\sim$4\,GHz.

To determine the expected $T_{\rm sys}$ we assumed an atmospheric contribution ($T_{\rm atm}$) of around 2\,K at these observing frequencies (\citealt{smi82}).  The contribution of ``spill-over'' radiation, $T_{\rm spill}$, from the ground entering the feed past the main reflector was estimated by integrating the feed response falling outside the reflector, assuming a ground temperature of 290\,K. This is shown in Figure~\ref{fg:tsys} as the green line. The relatively large variations are caused by frequency-dependent fluctuations in outer parts of the feed beam pattern. $T_{\rm scatt}$ is radiation from any source scattered into the feed from the telescope structure (our modelling does not account for scattering).  As far as possible, RFI is eliminated from the signal (see \S\ref{sec:rfi}) and so we do not consider it further in this section.  


The sky temperature depends upon the Galactic location being observed. An estimate of $T_{\rm Gal}$ was obtained from the reprocessed \cite{hssw82} 408\,MHz image of \cite{rdb+15} for the approximate sky position relevant to the measurements and assuming a spectral index of $-2.6$ (see range of values in \citealt{lmop87} for more details). Note that both the \cite{hssw82} and \cite{rdb+15} image scales include the cosmic microwave background contribution, $T_{\rm CMB}$, and so this was subtracted from the image value before scaling to the required frequency.   All these contributions have been plotted in the upper panel of Figure~\ref{fg:tsys} with the summation being shown as the upper, solid, black line.  

The measured on-telescope system temperature ($T_{\rm sys}$) as a function of frequency is shown in the top panel of Figure~\ref{fg:tsys} and tabulated in Table~\ref{tb:subbands}. These measurements were obtained by comparing spectra obtained with an ambient-temperature absorber over the feed with spectra obtained when observing the sky. The $T_{\rm sys}$  measurements were carried out with the telescope pointed at the zenith when both the Sun and the Galactic centre were below the horizon (between UTC 12:00 and 20:00 each day over the period 2018 October 29 to 31).  The $T_{\rm sys}$ values match, or significantly improve on, all current and previous Parkes receivers covering the same observing bands (see horizontal bars in Figure~\ref{fg:tsys}).  The predicted $T_{\rm sys}$ (from Equation~\ref{eq:tsys}) is within a few Kelvin of the measured value over most of the band, with slightly larger discrepancies at the low and high ends. At the low end, the discrepancy could easily be accounted by a small under-estimation of the spillover. At the high end, it may result from scattered power from various sources, including the telescope structure itself, entering the feed and/or leakage of ground radiation through the outer parts of the reflector.  We note that these $T_{\rm sys}$ measurements were determined from the output of the RF-chain and do not include any extra noise contributions from the digitiser and signal processor systems.

The black line in the bottom panel of Figure~\ref{fg:tsys} shows the expected aperture efficiency for the UWL as installed on the Parkes 64-m telescope. As described by \cite{sds+19} these were determined using laboratory-based measurements of the feed radiation patterns with a model of the telescope. The calculated efficiency is around 65\% over the observing band.  We compare these results with the astronomically-measured efficiencies described below. The upper, blue line in the bottom panel of Figure~\ref{fg:tsys} provides an estimate of the main beam efficiency. These values were also determined from the measured feed patterns using Equation~\ref{eqn:mainBeam}.  Across most of the band, the mean main beam efficiency is $\sim 0.96$, reducing to $\sim 0.92$ at the low-frequency end of the band.

\section{THE SIGNAL PROCESSING SYSTEMS}\label{sec:signalProcessor}

The two polarisation channels for each of the three bands are digitised in the focus cabin\footnote{The installation includes 12 analogue-to-digital converters (ADCs). However, only nine are used for the UWL system. The two polarisations of each of the three RF bands require six ADCs and a further three will be used in the future to provide a reference RFI signals from one or more reference antennas to be installed on the roof of the focus cabin. The remaining ADCs will be used for higher frequency receivers.}. The analogue-to-digital converters (ADCs) provide 12-bit resolution samples that are formatted into serial data streams complying with the JESD204B (Joint Electron Device Engineering Council Standard Document 204 Revision B) protocol for serially connected data converters. These signals are transported to the signal processors in the telescope tower via single-mode optical fibres, with two fibres required per digitiser\footnote{It is also possible to select specific frequency bands with up to 900\,MHz of bandwidth and pass those analogue signals through the Parkes down-conversion system and into the legacy signal-processor systems.}. The ADCs operate at one of two separate sampling frequencies, 2560\,MHz and 4096\,MHz, to ensure unbroken frequency coverage of the observing band from 704\,MHz to 4032\,MHz, as is shown in Figure~\ref{fg:samplebands}.

The six digitiser data streams (nine when the focus cabin RFI reference signal is implemented, see \S7) are fed to a pre-processor unit in the telescope tower which utilises Field Programmable Gate Arrays (FPGAs) to implement polyphase frequency channelisers that produce 26 contiguous sub-bands each 128 MHz wide for each polarisation (see Table~\ref{tb:subbands}). The pre-processor subsystem is built around a commercial platform with a Xilinx$^{\textregistered}$ Kintex$^{\textregistered}$ Ultrascale$^{\rm TM}$ FPGA and 8 Quad Small Form-factor Pluggable (QSFP) cages for data exchange. Each FPGA board receives data from two ADCs and eight such boards are housed in custom-made enclosures totalling eight rack units. The total designed capacity is 32\,GHz\ (256 channels of 128\,MHz) of real-time processing bandwidth and 1.28\,Tb/s of Ethernet data output. 

Currently, each sub-band delivered by the pre-processor is produced using a critically-sampled polyphase filterbank. To eliminate the effects of aliasing at the sub-band edges, an oversampled polyphase filterbank is planned, but has not yet been implemented.  There are two separate versions of the FPGA signal processing firmware to match the two ADC sampling frequencies. The primary differences between the two firmware versions are the number of output sub-bands and the length of the Finite Impulse Response (FIR) filter for the polyphase channelisers. For the 4096\,MHz sample clock firmware, there are 16 output sub-bands and 1024 coefficients in the channeliser filter. For the 2560\,MHz sample clock there are 10 output sub-bands and 640 coefficients in the channeliser filter. The channeliser FIR filters are carefully designed to achieve the best balance between pass-band flatness and attenuation of out-of-band components while also maintaining steep transition bands to minimise gain loss at the sub-band edges and aliasing between sub-bands. The FIR filters are designed using a least squares algorithm with a Kaiser window function to smooth the resulting impulse response, and have a pass-band ripple of approximately 0.02\,dB with a transition at the band-edges crossing the adjacent sub-band at $-$6\,dB. The relationship of the sub-band frequencies to ADC sampling frequencies is illustrated in the top panel of Figure~\ref{fg:samplebands} and tabulated in Table~\ref{tb:subbands}.

The polyphase filterbank is designed to preserve full numerical precision throughout both the FIR and Fast Fourier Transform (FFT) stages. Accordingly, there is ``bit-growth'' from the input 12 bits from the ADCs to 23 bits at the filterbank sub-band outputs. A number representation of 23 bits is not compatible with a CPU/GPU computational environment and would also require significant additional network infrastructure to transport. For these reasons, a scaling module is implemented for each sub-band output to reduce the number representation down to 16 bits (signed two's-complement). This scaling operation is an area where dynamic range issues of numerical overflow, underflow or saturation can occur, so the location of the 16\,bits that are selected from the input 23 can be controlled through software-adjustable fixed gain that has eight settings ranging exponentially from unity gain through to a gain of 128.

The output data streams from the FPGA pre-processor are sent as multicast Ethernet packets as 10\,Gb/s Ethernet (10 GbE) to a 64-port Cisco$^{\textregistered}$ 3164Q network switch (they are aggregated as 4$\times$10\,GbE on a 40\,GbE port on the switch).  During observations, the GPU-based Medusa signal processor subscribes to the multicast streams and runs further processing on them to form science data products, as described below. For the UWL RF bands, the FPGA signal processor outputs data as User Datagram Protocol (UDP) packets of size 8272\,B, each containing a single ``Data Frame'', as defined in the VLBI Data Interchange Format \citep{whi09} specification (VDIF; \url{http://www.vlbi.org/vdif/}). The FPGA pre-processor outputs a total of 52 VDIF UDP streams, one for each polarisation of each sub-band, with 128\,MHz bandwidth using complex sampling and 16-bit resolution per complex component; the total ingest data rate at the switch is 215\,Gb/s.

\begin{table*}
     \caption{Parameters for the 26 UWL sub-bands: the corresponding RF band that feeds them (column 2); sub-band frequency ranges (column 3); the central observing frequency (column 4); the median system temperature (column 5) and system-equivalent flux densities as determined using PKS~B0407$-$658 (column 6) summed over both polarisations; an estimate of the antenna (column 7) and main beam (column 8) efficiencies, the full-width-half-maximum (FWHM) of the measured beam shape (column 9), and the fraction of the sub-band containing useful data (column 10, see main text for definition).}
    \label{tb:subbands}
    \begin{center}
    \begin{tabular}{llcllllllll}
    \hline
         Sub- & RF& Freq. & Central & Median            & Median  & $\epsilon_{\rm ap}^{\rm sim}$  & $\epsilon_B$ & FWHM & Band \\
         band&band & range & Freq. & $T_{\rm sys}$ & $S_{\rm sys}$ & & & &  Use \\
          && (MHz) & (MHz) & (K)  &(Jy) &  & & (deg) & (\%) \\
         \hline
         0 &1& 704--832 & 768 & 39  & 49 & 0.70  & 0.92 & 0.43 & 31 \\
         1 &1& 832--960 & 896 & 31  & 45 & 0.69  & 0.95 & 0.38 & 24 \\
         2 &1& 960--1088 & 1024 & 22  & 38 & 0.66 & 0.96 & 0.35 & 65 \\
         3 &1& 1088--1216 & 1152 & 22  & 38 & 0.60 & 0.97 & 0.32 & 58 \\
         4 &1& 1216--1344 & 1280 & 22  & 38 & 0.64 & 0.96 & 0.27 & 68 \\ 
         \\
         5 &2& 1344--1472 & 1408 & 23  & 34 & 0.65 & 0.95 & 0.25 & 91 \\
         6 &2& 1472--1600 & 1536 & 23  & 36 & 0.67 & 0.95 & 0.23 & 79 \\
         7 &2& 1600--1728 & 1664 & 22  & 33 & 0.67 & 0.95 & 0.20 & 85 \\
         8 &2& 1728--1856 & 1792 & 22  & 34 & 0.66 & 0.97 & 0.20 & 67 \\
         9 &2& 1856--1984 & 1920 & 22  & 36 & 0.67 & 0.94 & 0.18 & 75 \\
         10 &2& 1984--2112 & 2048 & 22 & 36 & 0.65 & 0.96 & 0.18 & 92 \\
         11 &2& 2112--2240 & 2176 & 19  & 36 & 0.64 & 0.95 & 0.17 & 76 \\
         12 &2& 2240--2368 & 2304 & 21  & 39 & 0.62 & 0.96 & 0.18 & 62 \\
         \\
         13 &3& 2368--2496 & 2432 & 21  & 39 & 0.62 & 0.96 & 0.16 & 27 \\
         14 &3& 2496--2624 & 2560 & 21  & 39 & 0.61 & 0.95 & 0.16 & 92 \\
         15 &3& 2624--2752 & 2688 & 21  & 41 & 0.65 & 0.96 & 0.15 & 84 \\
         16 &3& 2752--2880 & 2816 & 20  & 39 & 0.64 & 0.96 & 0.14 & 92 \\
         17 &3& 2880--3008 & 2944 & 20  & 39 & 0.62 & 0.96 & 0.13 & 92 \\
         18 &3& 3008--3136 & 3072 & 21  & 39 & 0.64 & 0.95 & 0.13 & 91 \\
         19 &3& 3136--3264 & 3200 & 20  & 43* & 0.64 & 0.97 & 0.14 & 92 \\
         20 &3& 3264--3392 & 3328 & 20  & 45* & 0.62 & 0.96 & 0.13 & 92 \\
         21 &3& 3392--3520 & 3456 & 21  & 44* & 0.63 & 0.96 & 0.10 & 84 \\
         22 &3& 3520--3648 & 3584 & 23  & 48* & 0.63 & 0.96 & 0.12 & 79 \\
         23 &3& 3648--3776 & 3712 & 23  & 53* & 0.61 & 0.96 & 0.10 & 92 \\
         24 &3& 3776--3904 & 3840 & 25  & 70* & 0.60 & 0.96 & 0.11 & 91 \\
         25 &3& 3904--4032 & 3968 & 27  & 72* & 0.59 & 0.95 & 0.11 & 78 \\
         \hline
    \end{tabular}
    \end{center}
\end{table*}
 
The use of multicast Ethernet allows copies of the data to be sent not only to Medusa, but also to the Breakthrough Listen (BL) data recorder \citep{Price:2018}. Data is transported from the UWL switch to the BL switch via eight 40\,GbE links, which are configured as an aggregated group using Link Aggregration Control Protocol (LACP).

\subsection{The Medusa GPU cluster}

Medusa consists of nine rack-mounted, server-class machines, each equipped with dual Intel Xeon CPUs, 128 GB of random access memory, two 40\,Gb/s Network Interface Cards (NICs), four NVIDIA Titan X GPUs and an array of Solid State Disks (SSD). The 26 sub-bands are distributed across the nine Medusa servers, with each dual polarisation sub-band being received into a large, shared memory, PSRDADA (\url{http://psrdada.sourceforge.net/}) ring buffer. The sub-bands are then processed independently on the GPUs to form the desired astronomical data products. The \textsc{spip} (\url{http://github.com/ajameson/spip}) and \textsc{dspsr} \citep{vb11} software libraries are used to form the output data products for pulsar timing, pulsar searching or single-pulse studies, transient searching, spectral line and continuum studies and/or VLBI. These are written to the local file systems, and then transferred via a 40\,Gb/s NIC to the Euryale data-staging server.

For all astronomy modes and all sub-bands, Medusa first performs signal-processing tasks, such as  unpacking the VDIF data streams from the FPGA pre-processor and converting the signals to upper side-band (increasing sky frequency with channel number). Optional functions include the ability to adaptively mitigate RFI if a reference RFI data stream is available, and the formation of calibration spectra if a switched noise source is operating. After this initial processing, each 128-MHz sub-band is passed to the primary signal-processing system, which produces the requested astronomy data products.

Pulsar-related (search and fold mode) processing tasks have been developed around the \textsc{dspsr} software suite \citep{vb11}.  Coherent de-dispersion can be applied to both pulsar fold and search mode observations.  Such de-dispersion removes the intra-channel dispersive delays, but not the inter-channel delays.

The specific astronomical observing modes, which may be independently configured for each of the 26 sub-bands, are as follows:
\begin{itemize}
    \item {\bf Pulsar folding:} \textsc{dspsr} acquires the pulsar ephemerides from a copy of the ATNF Pulsar Catalogue \citep{mhth05} and user-generated pulsar parameters and uses the \textsc{tempo2} software package \citep{hem06} to predict the pulse topocentric period and phase. The voltage data are channelised (between 64 and 4096 channels in standard observing modes) with each channel coherently de-dispersed using a convolving filterbank algorithm. The data are then detected (the channelised voltages are converted to power) and polarisation products formed, folded into pulsar phase bins (between 8 and 4096) synchronously with the topocentric pulsar period and integrated to the desired sub-integration length (currently between 8 and 60 seconds). Either one, two or four polarisation products, respectively, AA$^*$+BB$^*$ (pseudo-Stokes I); AA$^*$, BB$^*$; and AA$^*$, BB$^*$, Real(A$^*$B), Imag(A$^*$B), where A and B are the two orthogonal polarisation signals and the $^*$ symbol indicates the complex conjugate, can be formed and recorded. The sub-integration data are then added to a PSRFITS \citep{hvm04} fold-mode file as 16\,bit values.  A typical observation in this mode would have 128 channels per sub-band, 1024 phase bins, 4 polarisation products and 30 second subintegration lengths.  This leads to a data rate of around 27\,MB per subintegration or $\sim$3\,GB for a 1\,hr observation.
    
    \item {\bf Pulsar searching:} The \textsc{dspsr} package includes \textsc{digifits}, which processes the sub-band voltage data to produce a set of PSRFITS search-mode data files. The data are channelised (8 to 4096 channels) and one, two or four polarisation products formed as for fold mode. These are averaged to a specified sampling interval ($\sim 1 \mu$s to 1\,s) and then they are rescaled to zero mean and unit variance using a computed scale and offset. These normalised samples are scaled and requantised to 1-, 2-, 4- or 8-bit values \citep{jen+98}. The offset and scale are retained for each block of samples (typically 4096 samples) and recorded with the time series data allowing for reconstruction of the original signal. The data volume and rate can vary enormously, depending upon the chosen parameters. For instance, for an observation with a sampling rate of 64$\mu$s, 1024 channels/sub-band, 2-bit sampling and 1-polarisation product (typical for recent surveys) the data rate would be 100\,MB/s or 370\,GB for a 1\,hr observation. However, to deal with dispersion smearing an observer may wish to observe with significantly narrower channel bandwidths for the low-frequency sub-bands, and wider bandwidths for the high-frequency sub-bands.   
    \item {\bf Pulsar single pulse studies and searches at a known dispersion measure (DM)}: if the DM of an observed pulsar is known, or if a pulsar search is being carried out with prior knowledge of the likely DM for any pulsar (or similarly, for searching for repeated fast-radio-burst events), then it is possible to coherently de-disperse the search-mode data stream and to produce PSRFITS search-mode data files. This allows for substantially reduced frequency resolution, which may then be traded for higher time resolution.  The data rate in this mode depends upon the chosen number of channels. If the pulsar DM is well known then, in theory, only a few channels are required for each sub-band, but having only a few channels can make RFI flagging challenging.  A typical observation mode may have 32$\mu$s sampling, 8-bit samples, four polarisation products and 128 channels for each sub-band. This produces 400\,MB/sec or 1.5\,TB in a one hour observation.
    \item {\bf Spectral line and continuum observations}: The data are channelised with up to $2^{21}$ = 2097152 channels per sub-band (equivalent to a frequency resolution of 61\,Hz) forming one, two or four polarisation products as for pulsar modes,  with a sampling interval between 0.25 and 60 seconds. These spectra are stored in Single Dish Hierarchical Data Format (SDHDF; see below). If available, calibration spectra may be stored in each output data file. Zoom bands will soon be supported by forming spectra spanning entire sub-bands at the desired frequency resolution and then discarding unwanted channels before the final data product is formed. One restriction of the UWL system is that the digitiser sampling frequencies are fixed. It is therefore not possible to calibrate the bandpass by frequency switching.  As above, the data rate can vary enormously, depending on the chosen parameters. With $2^{21}$ channels per sub-band the data rate is 436\,MB for each spectrum written to disk.  If one spectrum is produced each second then this corresponds to 1.6\,TB/hr.
\end{itemize}

 The switched noise source, used to provide a calibration signal, can be synchronised with the signal processing system with an observer-selected switching frequency.  Medusa can then produce separate calibrator-on and calibrator-off spectra, currently implemented for calibration of spectral data. Planned upgrades will enable the user to select the type of injected calibration signal and its frequency, phase and duty-cycle (see Section~\ref{sec:conclusion}). 

\subsection{Data staging and archiving}

The specifications for the data-staging server, Euryale, were defined so that it could receive the incoming data streams, manipulate the data files, interface with databases containing relevant information such as telescope pointing directions, have a disk buffer to store the data files and the ability to transfer the completed data files to the data archives. 

To optimise performance, the incoming data streams are balanced between two 40\,Gb/s network cards, four (non-volatile memory express; NVMe) disks (each $\sim 4$\,TB in size) and two Intel Xeon CPUs each with eight cores and two threads per core.   The data files can be large and the system needs to manipulate (including re-ordering and averaging) large data volumes so the Euryale server has 376\,GB of random access memory (RAM).  The server produces archive-ready data products by combining the data files from each of the sub-bands including calibration data, applying observational metadata, and writing out files in the required data format. The completed data products are temporarily stored on a large Redundant Array of Independent Disks (RAID) storage unit (this RAID provides a usable $\sim$86\,TB data store).  The data files are then sent to the relevant archives.

For the spectral line and continuum data sets we have developed SDHDF, a format based on the Hierarchical Data Format definition (HDF; \url{https://www.hdfgroup.org/}) initially used for the HIPSR system at Parkes for recording H{\sc i} observations using the 20\,cm multibeam receiver \citep{psb+16}. The SDHDF format is versatile and extendable, and able to be readily handled by modern computer languages. We are continuing to develop and define this data format and our final definition will be published elsewhere.   

All archive-ready data products need to include metadata for the observation. Euryale collects metadata from the observing systems and the GPU cluster and includes these values in the final data product. We normally ensure that no individual pulsar data file becomes larger than $\sim 10$\,GB by splitting observation data files in time and/or frequency.

Data sets for the majority of observations with the Parkes telescope become publicly available after an 18-month proprietary period.  The pulsar data sets are available from CSIRO's data archive (\url{https://data.csiro.au}; \citealt{hmm+11}). The spectral line and continuum data sets will be made available from the Australia Telescope Online Archive (ATOA; \url{https://atoa.atnf.csiro.au}) or
the CSIRO ASKAP Science Data Archive, CASDA \citep{cha17}.

\section{TIMING AND SYNCHRONISATION}\label{sec:timing}

An observatory distributed clock system, which is referenced to 
the observatory's hydrogen maser frequency standard, is used to derive precise 128\,MHz and 1\,Hz reference signals for the data acquistion systems. Phase-locked synchronisation signals at 2.56\,MHz for the 4096\,MHz system and 1.6\,MHz for the 2560\,MHz system, known as ``SYSREF'' signals, are required by the ADCs. These are derived from the 128\,MHz reference signals and distributed with it to the focus cabin over optical fibres. A local synthesiser in the cabin generates the sampling clocks from the 128\,MHz reference and distributes them to all ADCs along with a copy of the appropriate SYSREF clock. The combination of the sampling clock, SYSREF and a 1\,Hz gated synchronisation from the FPGA receivers, along with local copies of these signals supplied directly to the FPGA, allows for precise, repeatable synchronisation with deterministic latency. The ADCs are synchronised to a chosen SYSREF edge and therefore the synchronisation error is expected to be sub-nanosecond, relative to the maser standard. Time synchronisation to terrestrial time standards at a level of a few nanoseconds is obtained by measuring the offset between the observatory clock 1-second pulses and those from a Global Positioning System (GPS) receiver. These are subsequently referenced to International Atomic Time (TAI) or other time standards using time offsets published by the Bureau International des Poids et Mesures (BIPM)\footnote{https://www.bipm.org}.

\begin{figure}
    \includegraphics[width=5.8cm,angle=-90]{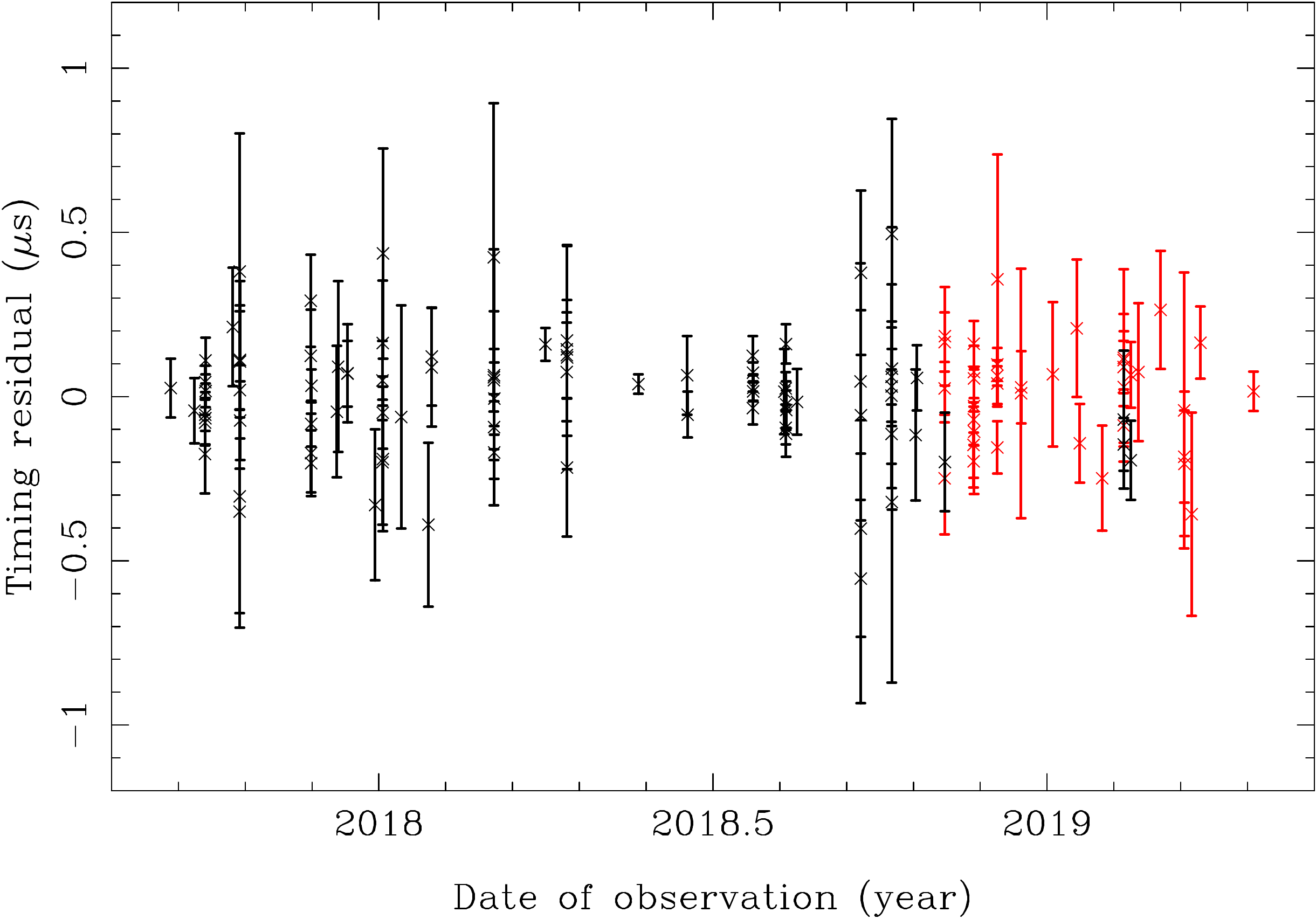}
    \caption{Timing residuals from PSR~J1909$-$3744 obtained by the Parkes Pulsar Timing Array project team. Points in black are from observations with the 10\,cm receiver and PDFB4 signal processor. Those in red are observations covering the same band, but with the UWL system.}
    \label{fg:timing1909}
\end{figure}

To illustrate the timing stability of the UWL system, Figure~\ref{fg:timing1909} shows the pulsar timing residuals for PSR~J1909$-$3744 in the 10\,cm observing band. This is the most stable pulsar observed by the Parkes Pulsar Timing Array project \citep{mhb+13}. The residuals in black were obtained using the 10\,cm receiver and the PDFB4 processor. Those in red are for an identical band, but obtained through the UWL system. The rms of the timing residuals (that span around 6 months) obtained with the Medusa system (over this relatively small band) is 145\,ns and the residuals are dominated by the uncertainty on their measurement.  The UWL is clearly adequate for high-precision pulsar timing, but we will continue to test the timing precision and accuracy of the system and to improve our knowledge of the delays between this system and earlier receiver and signal processor combinations.  Our intention is to determine the absolute delays required to convert any measured pulse arrival times to the intersection of the axes of the telescope (see \citealt{mhb+13} for details on how this was carried out for the PDFB4 signal processor).

\begin{figure*} 
\begin{center}
\includegraphics[width=11cm,angle=-90]{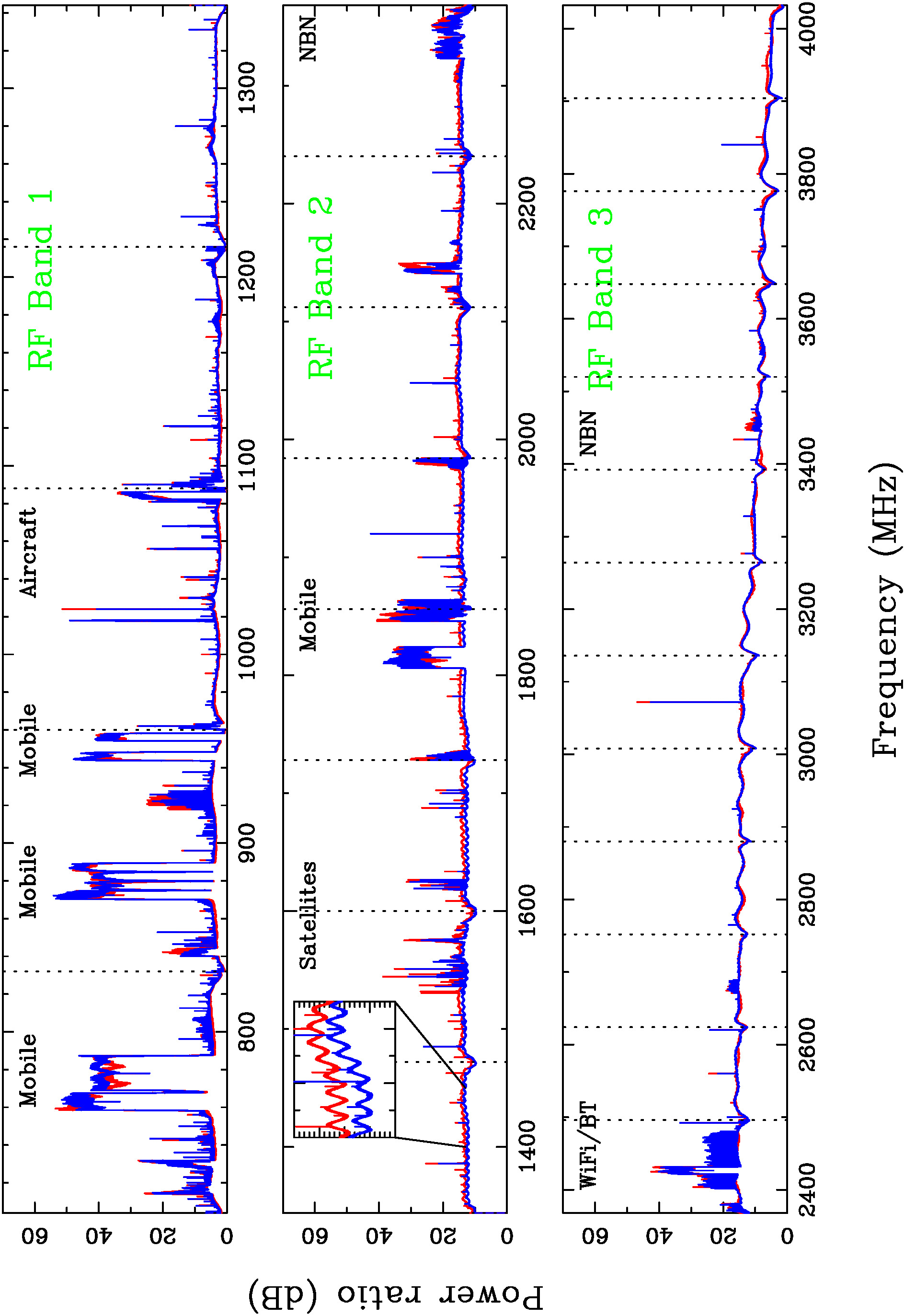}
\caption{Average spectra across three RF bands obtained on 2019 Feb 24 at 22:48:23 UTC. The red and blue traces represent the A and B polarisations respectively. The vertical, dotted lines indicate the sub-band boundaries and the principal RFI transmissions are labelled. The narrow spikes at 1024, 1920 and 3072\,MHz are digital processing artefacts related to the clock signals. NBN stands for the National Broadband Network and BT for Bluetooth. The zoomed-region in RF Band 2 highlights a characteristic ripple that is described in the text. The main text also provides an explanation of the slope across RF Band 3.}
\label{fg:spectra}
\end{center}
\end{figure*}


\section{SYSTEM PERFORMANCE}\label{sec:systemPerformance}

Here we describe the system performance as measured through the entire system (from the feed to the final data products).  Figure~\ref{fg:spectra} shows spectra with a frequency resolution of 488\,Hz across the three RF bands.  During this 2-minute observation the telescope was pointed towards the zenith.  The sub-band spectral shapes are significantly affected by RFI (see Section~\ref{sec:rfi}) and quasi-periodic oscillations.  
For instance the spectrum contains the characteristic small-scale ripple with a periodicity of $\sim 5.7$\,MHz (see zoomed-region in the Figure), which arises from reflections in the 26\,m space between the vertex and the underside of the focus cabin. 
The narrow-band spikes in the spectra are not yet fully understood; many will be externally generated RFI, although those at frequencies related to 1024 and 2560\,MHz are linked to the timing system and sampling frequencies.  Band 3 has a currently uncompensated $\sim$10\,dB slope across the band which originates in an RF amplifier within the digitiser module. This slope and the sub-band edges are stable and can be calibrated through observations by means of the injected calibration signal.

\begin{figure*}
    \centering
    \includegraphics[angle=-90,width=12cm]{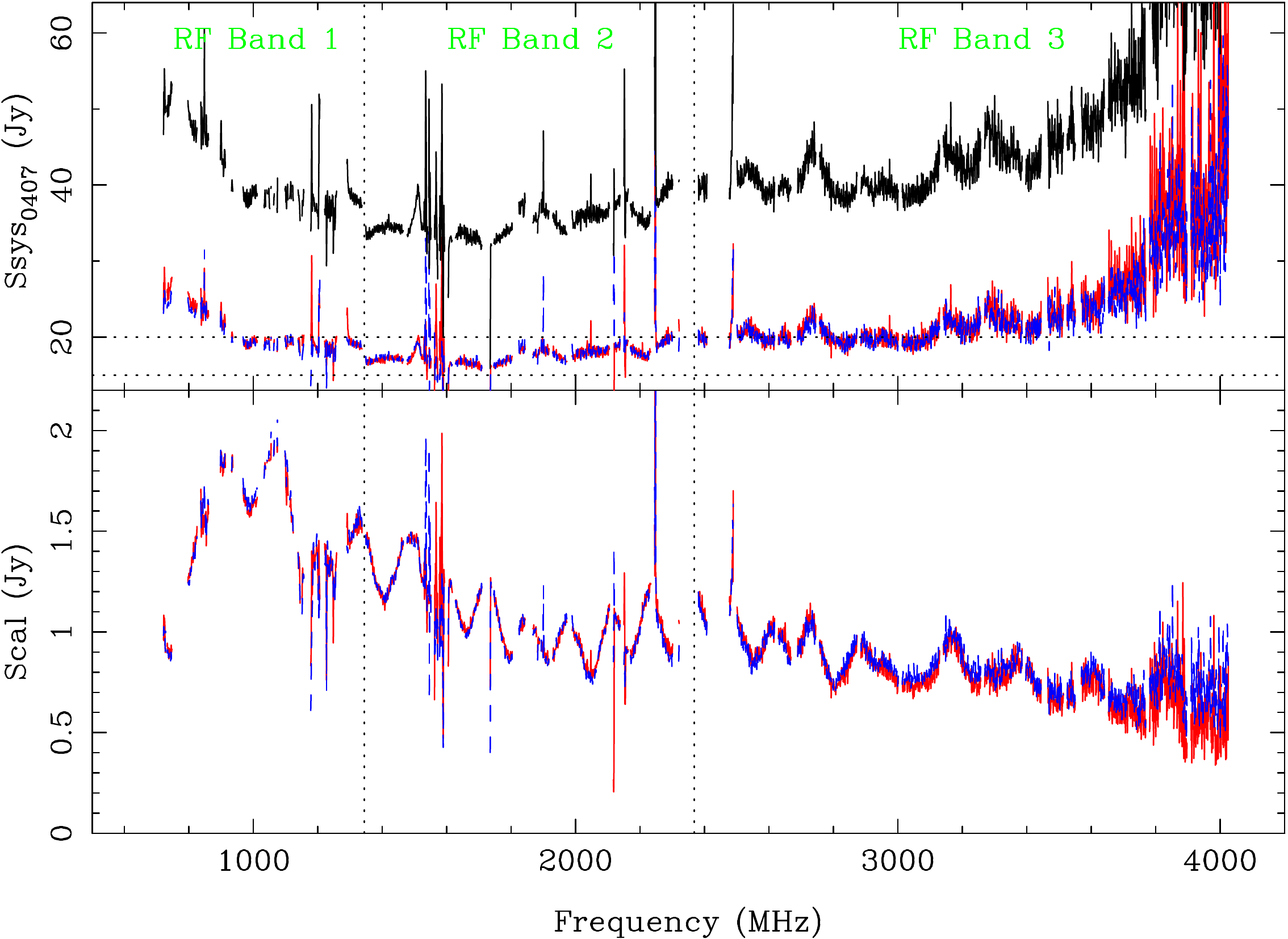}
    \caption{
    The measured system equivalent flux density (S$_{\rm sys}$) for the UWL receiver system on the Parkes radio telescope is shown in the upper panel, where the red and blue traces indicate the A and B polarisations respectively and the black line is the sum of the two polarisations. These measurements were obtained using observations of PKS~0407$-$658. The step between RF bands 1 and 2 is discussed in the text. The equivalent flux density of the injected calibration signal (S$_{\rm cal}$) for each polarisation is shown in the lower panel.}
    \label{fg:tsys_ssys}
\end{figure*}

The system equivalent flux density for the telescope with this receiver system ($S_{\rm sys}$) and the calibrator source ($S_{\rm cal}$) were determined by observing PKS~B0407$-$658 and its surroundings (we note that we obtain almost identical results with the calibrator PKS~B1934$-$638).  Each observation lasted two minutes at positions $1^\circ$ North of PKS~B0407$-$658, pointed directly at the source, $1^\circ$ to the South, again pointing at the source, and then again to the North. During each observation the noise injection is switched at a frequency of 11.123\,Hz.  The data were processed using the \textsc{Psrchive} program \textsc{fluxcal} to produce flux calibration files for each band. This process assumes that PKS~B0407$-$658 has a flux density of 15.4\,Jy at 1400\,MHz and a spectral index of $-$1.20 over the entire band (obtained from the Australia Telescope Compact Array calibrator database\footnote{ \url{http://www.narrabri.atnf.csiro.au/calibrators/calibrator_database.html}}).   The measured $S_{\rm sys}$ and $S_{\rm cal}$ values as a function of frequency are shown in the upper and lower panels of Figure~\ref{fg:tsys_ssys} respectively.  The lower red and blue lines in each panel of the Figure show the determinations independently for each polarisation. The upper, black line in the top panel is the more commonly quoted summation of both polarisations.  The median $S_{\rm sys}$ values for each sub-band are listed in Table~\ref{tb:subbands}.

Whereas the $T_{\rm sys}$ measurements, described in \S\ref{sec:receiver}, were obtained from the output of the receiver system whilst the telescope was pointing at zenith, the $S_{\rm sys}$ measurements have been determined using an astronomical source through the entire UWL system (including the digitisers, FPGAs and GPU processors). The discontinuous break in the $S_{\rm sys}$ measurements seen between RF bands 1 and 2 in Figure~\ref{fg:tsys_ssys} results from the extra attenuation required in Band 1 to avoid saturation effects in the analogue or digital systems which could otherwise occur because of the very strong mobile-phone transmitters in this band\footnote{For this Figure, the flexible attenuation levels were set to 15, 10 and 7\,dB in RF bands 1, 2 and 3 respectively.}. As Figure~\ref{fg:spectra} shows, these transmitters can be up to 50~dB above the receiver noise floor. However, this extra attenuation results in a significant digitiser noise contribution, adding an extra $\sim 3$\,Jy to the measured system equivalent flux density.  

The $S_{\rm cal}$ values clearly contain an oscillation with a $\sim 100$\,MHz periodicity. The noise source is injected into the LNA system, but the coupler is not directional. This oscillation is believed to result from reflection off the tip of the dielectric spear.  The oscillation period and structure are stable and, to date, have not affected the  calibration of astronomical data sets.

Using Equation~\ref{eqn:eap} in Appendix~A, we can obtain estimates for the aperture efficiency, $\epsilon_{\rm ap}$, from the system equivalent flux density values and the system temperature values (Figure~\ref{fg:tsys_ssys}). We note that the T$_{\rm sys}$ measurements were obtained at nighttime, with the telescope pointing at zenith using an independent spectrum analyser and hence do not include extra noise from the digitisers.  However, the $S_{\rm sys}$ measurements were obtained with a telescope zenith angle of 40$^\circ$ (elevation angle of 50$^\circ$). We therefore have extra contributions to T$_{\rm sys}$ from the atmosphere, spill-over and from the digitiser system.  For a typical sub-band (we have chosen sub-band 5 spanning from 1344 to 1472\,MHz) we estimate an extra contribution to T$_{\rm sys}$ during the PKS~0407$-$658 observations of $\sim 4$\,K giving $\epsilon_{\rm ap} \approx 0.7$.  From Equation~\ref{eqn:gain} in Appendix~A this converts to a gain of $G_{\rm DPFU} \sim 0.8$\,K\,Jy$^{-1}$.

\begin{figure*}
    \centering
    \includegraphics[angle=-90,width=8cm]{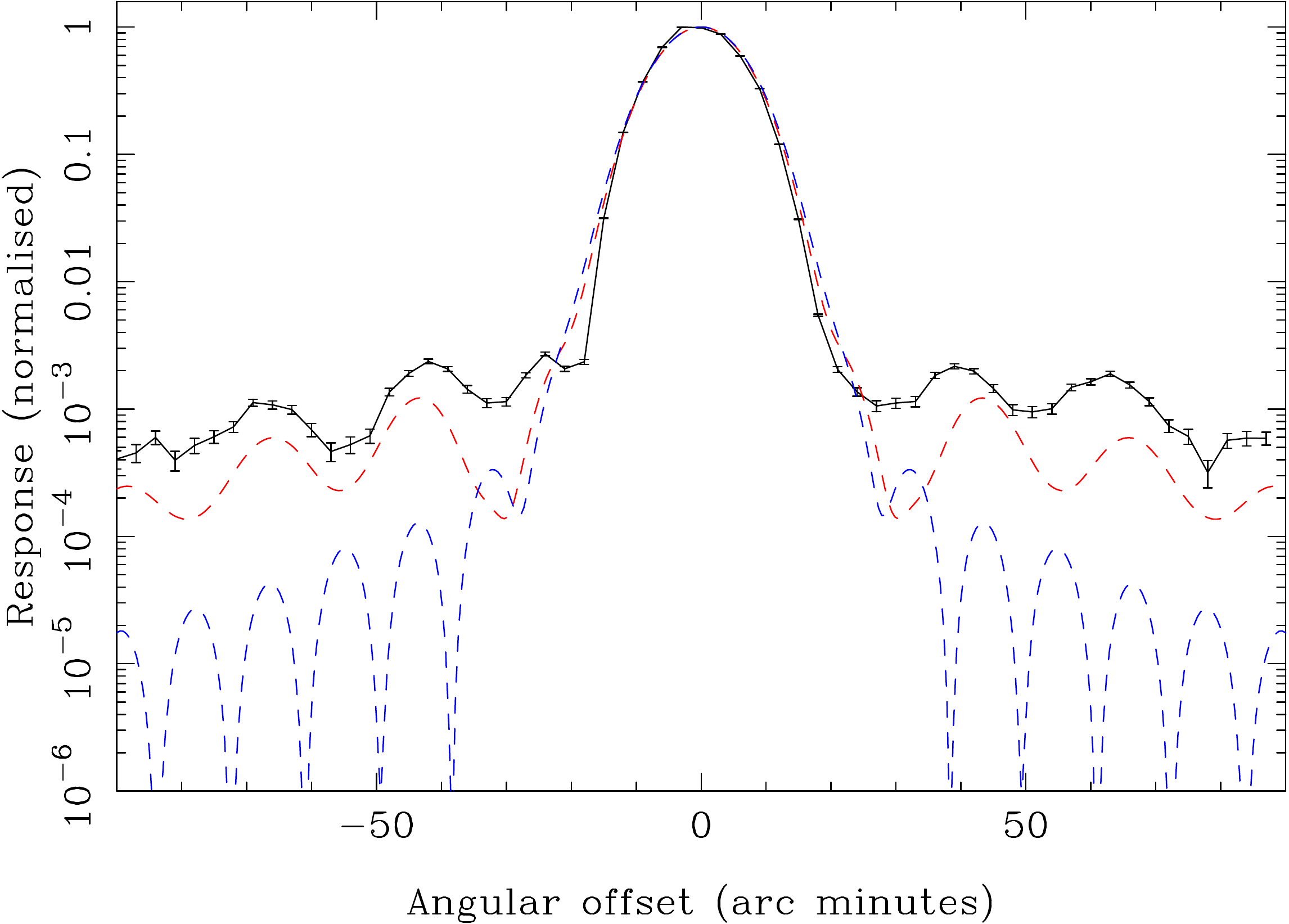}
    \includegraphics[angle=-90,width=8cm]{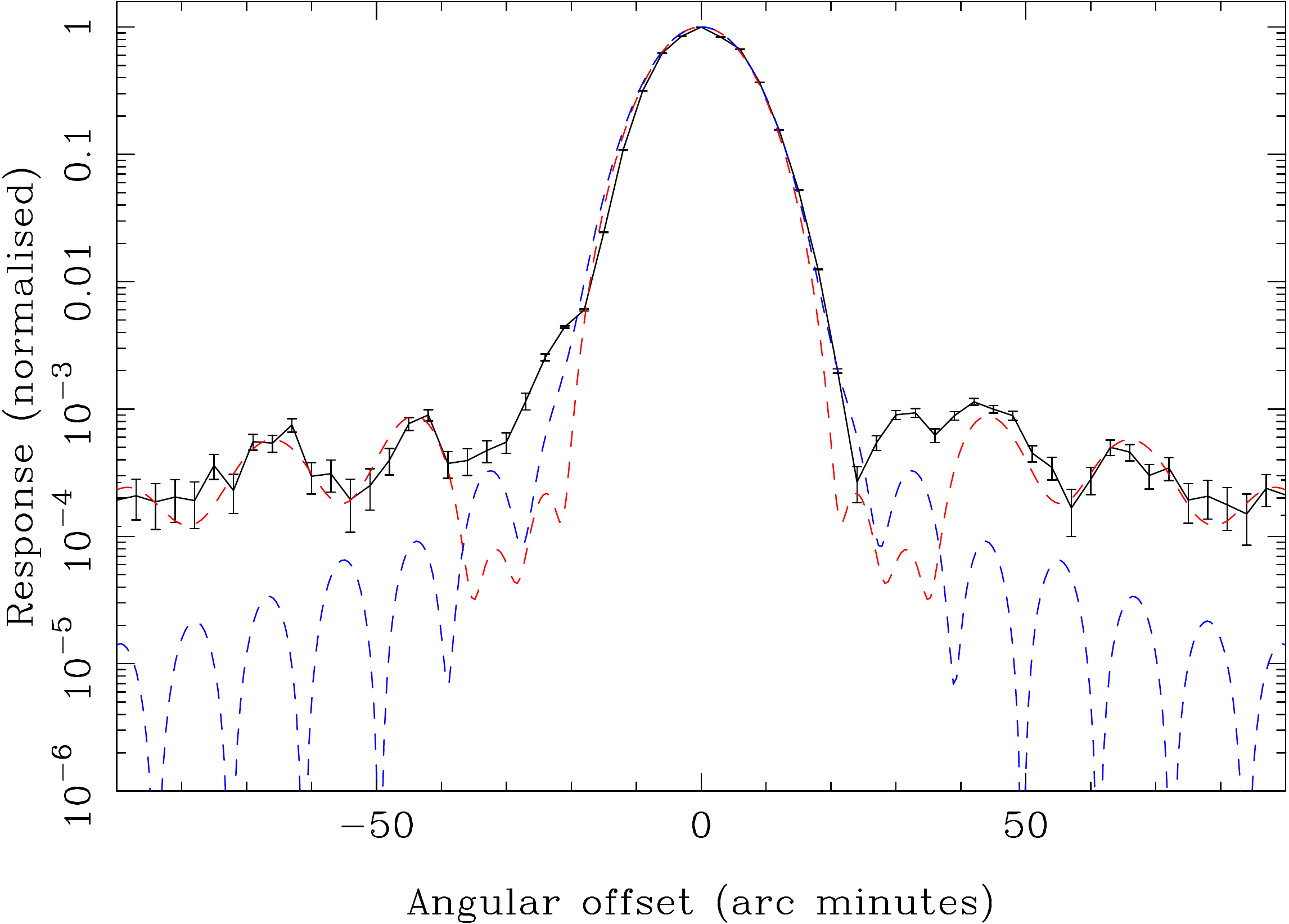}
    \caption{Beam pattern obtained through scans in azimuth (left)  and elevation (right) for the 20\,cm observing band (sub-band 5, Table \ref{tb:subbands}). The measured response is shown using the black line and error bars.  The red and blue dashed lines are modelled beam shapes that are described in the text.}
    \label{fg:beam}
\end{figure*}

The feed has been designed to have a constant beam width across the band \citep{sds+19}. However, the telescope beam width as measured on the sky remains proportional to the observing wavelength.  Theoretical full-width half power beam widths calculated from $1.02\lambda/D$ (where $\lambda$ is the observing wavelength and $D$, the telescope diameter) give 23, 12 and 4 arc minutes for 700, 1400 and 4032\,MHz respectively.  A high dynamic-range measurement of the actual beam shape over the wide-band can be obtained by determining the flux density (Stokes I)  of the Vela pulsar (PSR~J0835$-$4510) as a function of the angular separation between the known pulsar position and the telescope pointing direction (we define the pointing in azimuth, $A$, and elevation, $E$).  We have carried out multiple observations of Vela with the telescope offset in a grid of ($A \cos E$, $E$) positions. Figure~\ref{fg:beam} shows representative beam shapes (for sub-band 5 and centred on 1408\,MHz).  The complete set of beam shape data files is available for download (see Appendix B). The results from the grid pointings in elevation and in azimuth are not identical. This is primarily caused by the presence of a feed leg in the elevation plane.  We overlay predictions of the beam shape obtained from electromagnetic simulations.  The blue dashed lines are an ideal case with no attempt to model the effect of the prime-focus cabin.  The red, dashed line includes a simplified model of the cabin (assumed to be a circular blockage), but does not include the effect of the feed cabin support legs.  We do not expect a perfect match between the observing beam shape and the prediction, but note that they match remarkably well and that the side-lobe structure is dominated by the effect of the focus cabin.  Data files containing the predictions are also available as part of our public data collection.

The on-axis polarimetric response of the receiver system is modelled using an updated version of the Measurement Equation Modeling (MEM) technique originally described in \cite{van04}. The required updates, motivated by \cite{lck+16}, remove the assumption that the system noise has insignificant circular polarisation and enable modelling of an artificial noise source that is coupled after the feed.  
We have used this updated MEM algorithm to measure the properties of the receiver (including the differential gain, differential phase, and cross-coupling of the receptors; and the Stokes parameters of the artificial calibration signal). 
As inputs, we observed the bright millisecond pulsar, PSR~J0437$-$4715, over a wide range of parallactic angles; we also included both on-source and off-source observations of Hydra~A (3C218).
The measured receptor ellipticities are close to $0^\circ$ across the whole band, indicating that the degree of mixing between linear and circular polarisation is low. 
The non-orthogonality of the receptors is also very low, as characterised by the intrinsic cross-polarisation ratio \cite[IXR;][]{cw11}, which varies between 40 and 60\,dB across the band. This is much greater (noting that larger values of the intrinsic cross-polarisation ratio correspond to lower non-orthonormality) than the minimum recommended by \cite{fkp+15} of 30\,dB for high-precision pulsar timing.
We also found that, at higher frequencies, the reference signal produced by the artificial noise source deviates from 100\% linearly polarised. Up to $\sim30$\% circular polarisation has been observed at $\sim$4\,GHz; therefore the polarisation of astronomical signals must be calibrated using the technique described in Section 2.1 of \cite{ovhb04}.

We provide the set of calibrator parameters as determined using the updated MEM algorithm as part of our publicly-available data collection (see Appendix~B).

We have not yet studied the off-axis polarisation properties of the system in detail, but provide theoretical models for the off-axis polarisation purity in our data collection.

\subsection{Demonstration of astronomical observation modes}

\begin{figure*}
\begin{center}
\includegraphics[width=16cm,trim={1.5cm 3.3cm 1.0cm 1.5cm},clip]{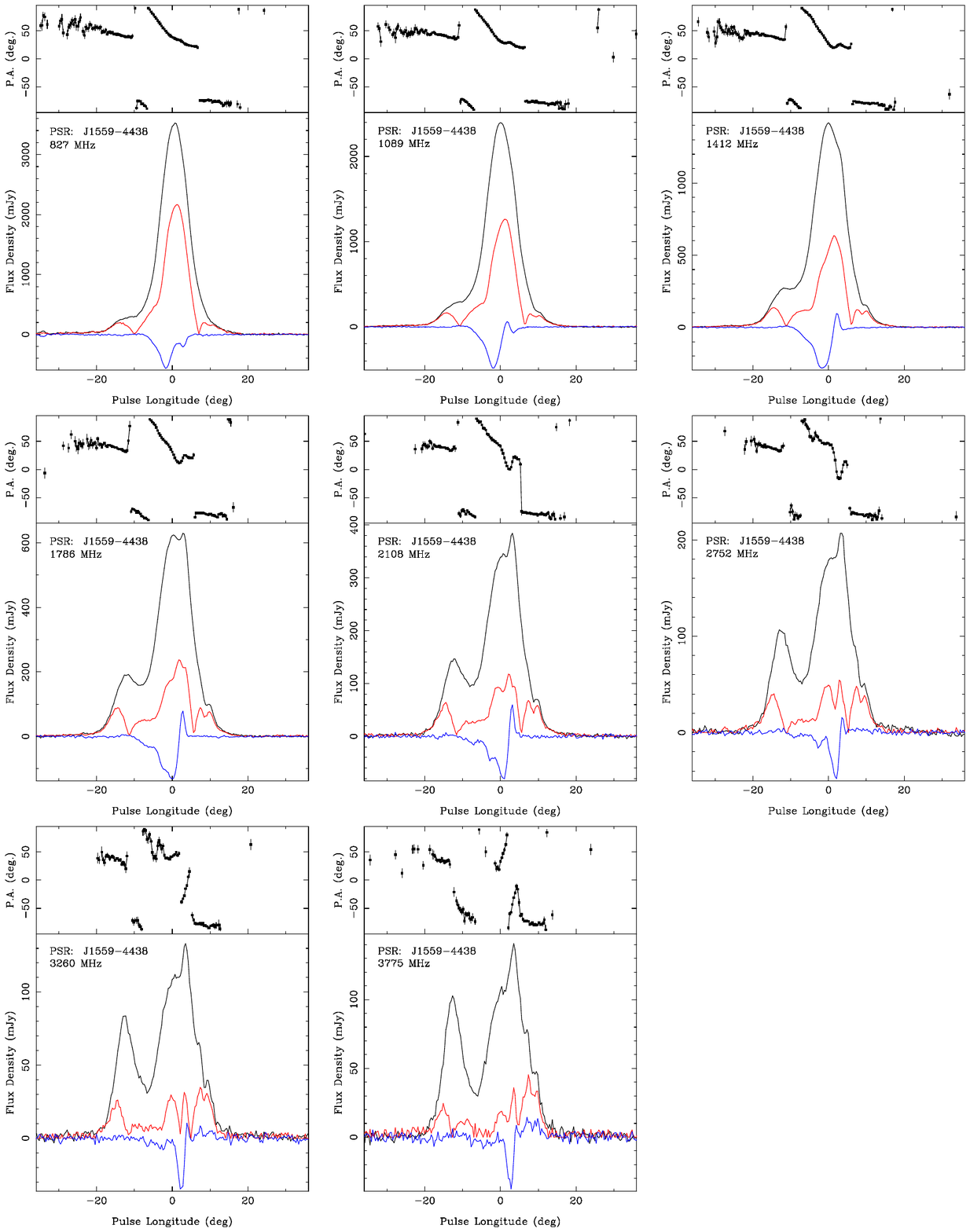}
\caption{Pulse profiles for PSR~J1559$-$4438 (B1556$-$44) in eight different frequency bands. The top section of each panel shows the position angle of the linear polarisation. The profiles are shown as total intensity (black), linear polarisation (red) and circular polarisation (blue). The position angles (P.A.) have been corrected in infinite frequency assuming a rotation measure of $-5$\,rad\,m$^{-2}$.}\label{fg:pulsar}
\end{center}
\end{figure*}

 To demonstrate the end-to-end system we show, in Figure~\ref{fg:pulsar}, a wide-bandwidth observation of PSR~J1559$-$4438 (B1556$-$44). The observation has been divided into eight frequency channels.  The observation has been both flux density and polarisation calibrated. Details of the pulse profile variations with frequency will be published elsewhere, but we note that these results agree, in common observing bands, with previous measurements (e.g., \citealt{jkm+08} and \citealt{jk18}).

\begin{figure}
    \centering
    \includegraphics[width=5.5cm,angle=-90]{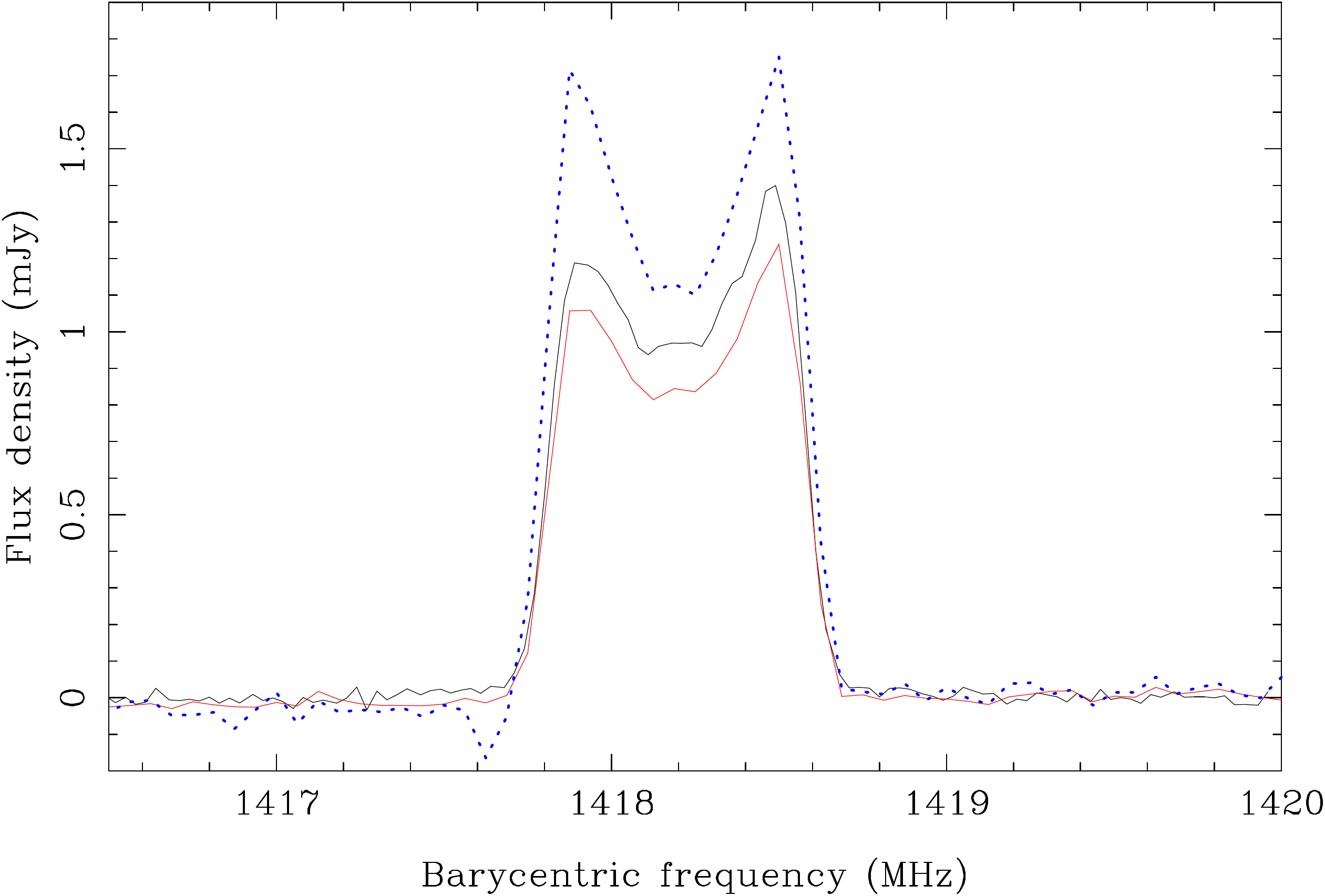}
    \caption{The UWL spectrum of the H{\sc i} emission from NGC 45 (black) overlaid on the expected HIPASS spectrum (red).  The UWL and HIPASS spectra have spectral resolutions of 31.25\,kHz and 62.5\,kHz and integration times of 300 and 450\,s respectively. The, higher, blue-dotted line is a reprocessing of the HIPASS data cube using the Duchamp source finder and integrating over the entire extended structure of the source.}
    \label{fg:spectrum}
\end{figure}

An example of the spectral-line observing mode is shown in Figure~\ref{fg:spectrum}. We observed both NGC~45 (HIPASS~J0014$-$23) and an off-source position for 5 minutes. The off source position was 1.2 degrees offset in both azimuth and elevation. 
We recorded 4096 channels per sub-band and a spectrum was written to disk every second. During both observations the calibration signal was switched at 100\,Hz and the output data product (available as part of our public data collection) contains lower resolution spectra representing the on and off states of the calibrator as well as the astronomy spectrum.  The calibration process for the astronomy spectrum was carried out offline, using the on/off calibration spectra and the on/off source observations.  The resulting H{\sc i} spectrum (black) is overlaid on the spectrum obtained from the HIPASS survey \citep{bsb+01} (red line). We note that this source is slightly extended and so we also overlay in the Figure (blue, dotted line) the integrated spectrum obtained after identifying the source in the original HIPASS data cube (H236) using the Duchamp source finder \citep{whi12}. The spectral shape matches well with the HIPASS result and our flux scale lies between the HIPASS spectrum and Duchamp results.  

\section{RADIO FREQUENCY INTERFERENCE}\label{sec:rfi}

All radio-astronomy receiver systems are affected by RFI.  Ground-based, aircraft and satellite transmissions are especially problematic for wideband centimetre-wavelength systems such as the UWL.  Figure~\ref{fg:spectra} shows the entire UWL spectrum, with some known RFI sources labelled.  

In Australia, licensed ground-based transmitters can be identified through the Australian Communications and Media Authority (ACMA) website\footnote{\url{https://www.acma.gov.au}}. These include numerous mobile phone signals including the 3rd generation (3G) networks between 850 and 900\,MHz and the 4th generation (4G) networks around 760, 780, 1800, 2100 and 2680\,MHz. National Broadband Network (NBN) transmissions exist around 2.37\,GHz and 3.56\,GHz.

As Figure~\ref{fg:spectra} shows, the strongest persistent RFI bands are mobile-phone transmissions in RF Band 1. These are primarily from transmission towers located 10 to 20\,km away, both north and south of the telescope. These RFI signals enter the system through far-out sidelobes of the antenna response and hence are quite variable depending on antenna pointing. Despite the low gain of the antenna sidelobes, these signals can be 50\,dB or more above the local noise floor. Over the full RF Band 1 bandwidth, they typically add up to 35\,dB of power to the digitised signal.  Additional attenuation is required in the RF Band 1 chain to avoid saturation effects and the consequent formation of inter-modulation products across the RF Band 1 spectrum. RF bands 2 and 3 do not suffer from these problems since the RFI signals are much weaker in these bands.

\begin{figure*}
\begin{center}
    \includegraphics[width=12cm,angle=-90]{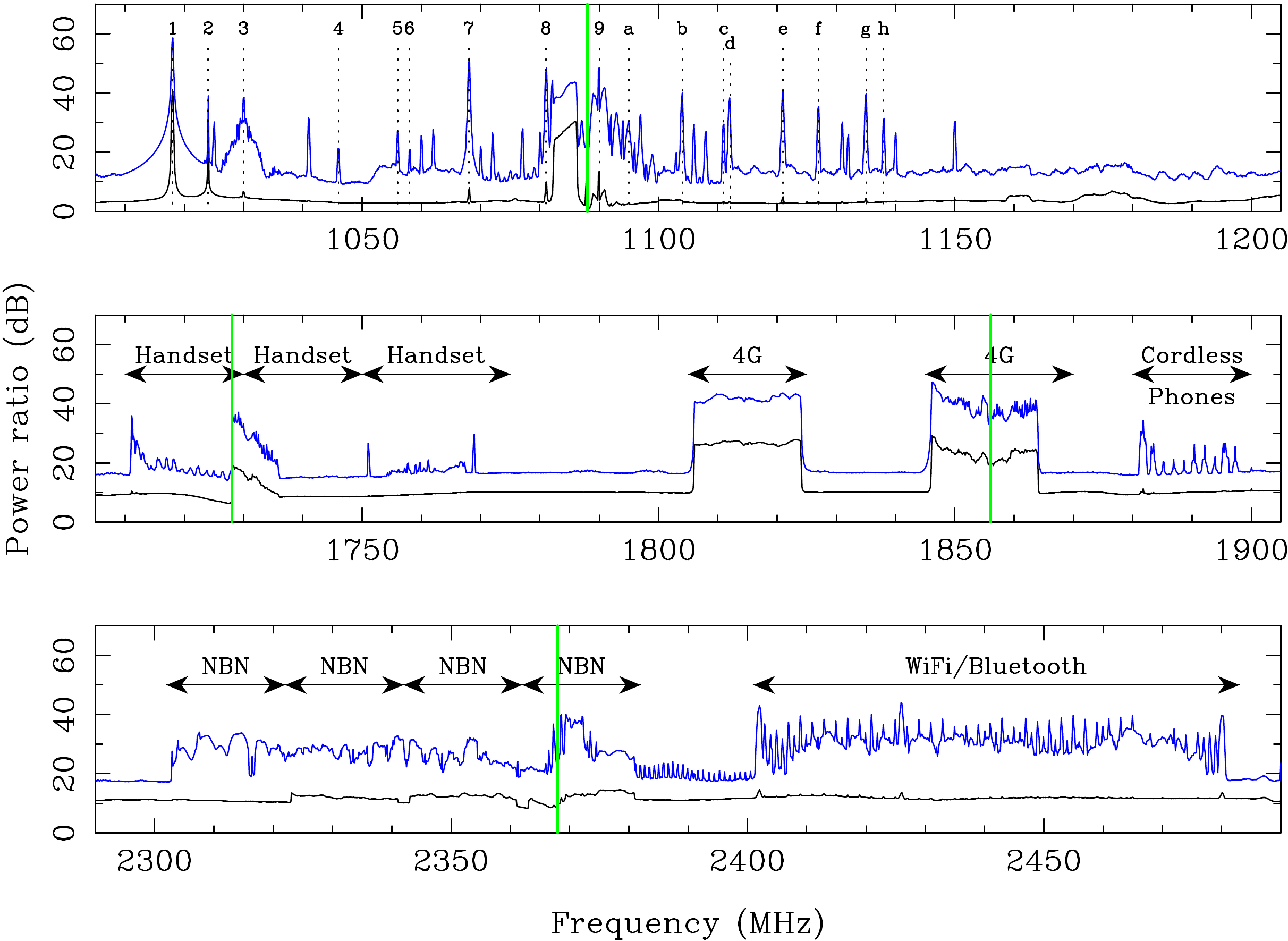}
    \caption{Mean (black line) and peak-hold (blue line) spectra for parts of the band that contain significant impulsive interference. The vertical, green lines indicate sub-band boundaries. Labels in the upper panel indicate DME signals from the following towns and cities: Parkes (1 and 8), Nowra (4 and g), Canberra (5 and h), Sydney (6 and d), Richmond (7), Cowra (a), Williamtown (b), Cooma (c), Wagga Wagga (e) and Albury (f). ADS-B is indicated by 9 and air traffic control ground interrogation by 3. An internal clock system produces signal 2 at 1024\,MHz. The broad-signature around 1730\,MHz is an alias of strong RFI at the 1860\,MHz and the broad signal around 1085\,MHz is an alias from a strong mobile-phone band just below 960\,MHz. Several of the strong peaks, especially for the peak-hold data in the top panel, have wide sidebands; these originate from the short FFTs used to provide the high time-resolution data.}
    \label{fg:peakHold}
    \end{center}
\end{figure*}

There are also numerous satellites emitting in the UWL band including the various global navigational satellite systems (such as GPS, Beidou, GLONASS and Galileo) as well as communication systems such as the Thuraya (1525 to 1559\,MHz) and Iridium (1617 to 1627\,MHz) satellites.  The positions of satellites for which public ephemerides are available can be monitored by the observer during an observation. For post-observation analysis of suspect or distorted bandpasses, an online resource exists\footnote{\url{https://www.narrabri.atnf.csiro.au/cgi-bin/obstools/sarfis.pl}} permitting the observer to find RFI generating satellites that might have been near the main beam of the telescope at a given time in the past.

Short duration impulsive or transient RFI signals are also common.  Because of their typically low duty cycle they are not prominent in spectral plots such as Figure~\ref{fg:spectra}, but can affect astronomical observations. There are two main classes for such signals: a) broad-band signals, mainly from switching transients in local electrical systems or from lightning, and b) highly modulated relatively narrow-band (typically a few MHz) transmissions, from a variety of sources, for example, aircraft navigation systems, mobile phone handsets and many satellite transmissions.  

On June 5, 2019, we observed using a modified version of the spectral line/continuum mode that recorded maximum signal powers in 64\,$\mu$s samples of each 125\,kHz channel, as well as providing the average power in each channel. The observation lasted for 15 minutes. 
In Figure~\ref{fg:peakHold} we have selected three regions that exhibit significant transient RFI.  The black line in this Figure gives the mean spectrum and the blue line indicates the peak sample values recorded for each channel in the 15\,min observation.

Aircraft navigation signals cover a wide band from 962 to 1150\,MHz and typically consist of sequences of microsecond-duration pulses with a low duty cycle. Consequently, they contain very little power on average but are prominent in the peak-hold spectra.  The airport in Parkes has a Distance Measuring Equipment (DME) beacon transmitting at 1018\,MHz with corresponding airborne beacons interrogating on 1081\,MHz; both are strong signals in the UWL spectrum. Numerous other signals from other DME systems are seen between 1020 and 1150\,MHz. 

Aircraft-borne Automatic Dependent Surveillance - Broadcast (ADS-B) systems transmit at 1090\,MHz. In addition, ground-based air traffic control radars send an interrogation signal at 1030\,MHz. The 1090\,MHz signals will be visible at ranges of up to hundreds of kilometres for aircraft at cruising altitude and so at any given time signals from many aircraft will be seen. ADS-B transmissions consist of short sequences (lasting $\sim 120$\,$\mu$s) of  digitally-encoded bursts with low duty cycle, where each individual burst is only 0.5$\mu$s in duration.

We know that much of the broad-band impulsive interference that we detect is locally generated.  There were initial concerns that our focus cabin digitisation system would generate an unacceptable level of RFI. However, we estimate that the digitiser packaging and the shielded cabinets together provide at least 100\,dB of RFI attenuation and no such leakage has been detected in the recorded signals. Even though active mobile, Wi-Fi and/or Bluetooth devices are not permitted on site, we do observe signals across these bands. We are still investigating the source of these signals, but it is likely that some originate in electronic devices used by the general public at the Telescope Visitors Centre, which is only $\sim 100$\,m from the telescope.

In Table~\ref{tb:subbands} we list the fraction of each sub-band that was removed by automatic pulsar data processing scripts in more than 50\% of 236 pulsar observations taken on 2019 February 17.   In total, such automated processing removes $\sim$814\,MHz of the band.  To reject aliasing effects resulting from the use of critically sampled digital filterbanks we remove $\pm 5$\,MHz at each sub-band boundary, a total of 260\,MHz of bandwidth. The planned implementation of over-sampled filterbanks will remove the need to reject these sub-band edges.

\section{FUTURE DEVELOPMENTS AND CONCLUSION}\label{sec:conclusion}

Iterative improvements will continue to enhance the UWL receiver system. Planned upgrades include a novel, single-dish calibration scheme in which the noise source is modulated in a pseudo-random sequence continuously throughout an observation.  The signal-processor system can then use the known pseudo-random sequence to extract the time-dependent calibration solutions.

A further planned upgrade will be to develop over-sampling (replacing critical-sampling) of the input data stream, preventing signal degradation at the sub-band edges and effectively eliminating aliasing effects. This will require updates to the FPGA firmware.

The current UWL system provides a platform to research RFI mitigation techniques. Transmissions from mobile phone and NBN towers are confined in bandwidth and are quasi-continuous. In principle such signals can be removed using adaptive filtering techniques based on having an RFI reference signal that can be correlated with the astronomy signal \citep{kmbh10}. In practice, this is difficult because of the great strength of the interfering signals, requiring very high S/N in the reference signal, and the fact that multiple sources contribute to the signal in some RFI bands. For example, the 4G band containing 760\,MHz is transmitted from both Alectown, about 11\,km north of the telescope, and from a tower near Parkes, about 15\,km south-west of the telescope.   In principle, some satellite signals could also be removed using adaptive filter techniques, but obtaining a reference signal with sufficient S/N is a challenge.  Signals that are easily detectable in the peak-hold spectra, but only marginally visible in the average spectra, are best removed in the time domain. Signal processing algorithms to achieve this are currently being evaluated but are not yet implemented in the standard pipeline processing  (for instance, \citealt{ng10}). Localisation and removal of  local sources of transient RFI remains an on-going effort. 

The telescope control system is currently being upgraded to a web-based system. When complete this will enable a range of observing modes, from almost hands-off scheduled observing, to hands-on control of the telescope and observing system. 

The astronomical signal-processing system will also be upgraded to provide (1) commensal observing modes, (2) transient detection capability, (3) the ability to fold multiple pulsars simultaneously, (4) to phase resolve spectra within a pulsar period and much more.  We will continue to characterise the system including using holographic measurements to study beam patterns in detail \citep{igs+19}. 

The observing system described here is provided to the astronomical community.  Teams can propose to observe with the system twice per year using the standard Australia Telescope National Facility proposal system (\url{https://opal.atnf.csiro.au/}). One of the primary challenges facing users of the system is the massive data volumes. Observation data files can relatively easily become TBs in size. Not only are such files difficult to transfer to local computer systems, our current astronomical software suites for single-dish telescopes are ill-equipped to deal with these data volumes. In addition, traditional calibration procedures such as feed rotation and frequency-switching cannot be applied and new methods will be required to extract the required scientific results from the wide-band data (such as new wide-band timing methods; e.g., \citealt{pen19}).

In the longer term we plan to upgrade the Parkes receiver suite with higher frequency, single-pixel wide-band receivers and a cryogenically cooled, phased-array-feed.  With this suite of receiver systems and versatile signal-processor and control systems, the Parkes telescope will continue to provide cutting edge observations of astronomical sources in the Southern hemisphere for many years to come.

\begin{acknowledgements}

The ultra-wide-band receiver project was primarily funded through an Australian Research Council Linkage Infrastructure, Equipment and Facilities (LIEF) grant, with additional funding obtained from the Commonwealth Scientific and Industrial Research Organisation (CSIRO), the Max-Planck-Institut f\"{u}r Radio Astronomie (MPIfR) and National Astronomical Observatories of the Chinese Academy of Sciences (NAOC).  We thank the Parkes Pulsar Timing Array team for providing UWL observations of PSR~J1909$-$3744 and Hydra A.  Staff at each of the institutions that have made contributions toward this project are acknowledged. 
We thank everyone who has provided advice, updated their software packages, or helped build, install and/or commission the system.  We thank Simon Hoyle for his work in developing the calibration system, Robert Shaw and Simon Mackay for their involvement in the front-end system and Euan Troop for software systems. We acknowledge B. Gaensler, S. Wyithe, Y. Levin and A. Melatos who were involved in obtaining the grant funding, but chose not to be an author on this paper. MB and SO acknowledge Australian Research Council grant FL150100148. Parts of this research were conducted by the Australian Research Council Centre of Excellence for Gravitational Wave Discovery (OzGrav), through project number CE170100004.

The Parkes radio telescope is part of the Australia Telescope National Facility which is funded by the Australian Government for operation as a National Facility managed by CSIRO.
\end{acknowledgements}

\begin{appendix}

\section{Definitions}

Radio engineers \citep[e.g.,][]{kra66,stu98} recognise a dimensionless gain $G$ or, equivalently for a lossless antenna, a directivity $D$ which expresses the power transmitted from, or received by, an antenna in a particular direction with respect to an idealised antenna that radiates or receives power isotropically. Antenna gain so defined is frequently specified logarithmically in dB$_{\rm i}$ (``decibels-isotropic''). An alternative metric often used in radio astronomy is the effective area $A_{\rm e}$ of an antenna, relating the power density $p$ (in W\,Hz$^{\rm -1}$) received by the antenna from a radio source of flux density $S$ (in W\,m$^{\rm -2}$\,Hz$^{\rm -1}$) through the defining relation:
\begin{equation}
    p = A_{\rm e} S
\end{equation}
where the radio source and antenna are assumed to have matched polarizations and $A_{\rm e}$, like its counterpart $G$, is in general a function of frequency, direction and polarization. As large radio telescopes typically possess an obvious physical or geometric aperture it is natural to define an aperture efficiency $\epsilon_{\rm ap}$, as the ratio of the effective to the physical collecting area:
\begin{equation}
    \epsilon_{\rm ap} \equiv \frac{A_e}{A_p}
\end{equation}
where $A_{\rm p}$ is the physical (geometric) aperture of the telescope.
It can be shown that $G$ and $A_{\rm e}$ are related thus:
\begin{equation}
    G = \frac{4\pi}{\lambda^2}A_{\rm e} = \frac{4\pi}{\lambda^2}\epsilon_{\rm ap}A_{\rm p}
\end{equation}
where $\lambda$ is the observing wavelength.

Another proxy for antenna gain arises from the concept of antenna temperature $T_{\rm a}$, defined as the temperature of a fictitious matched resistor which, if substituted for the antenna, would produce the same increase in power density at the input to the receiver as an unpolarized radio source of flux density $S$. It can be shown (e.g., \citealt{kra66}) that:
\begin{equation}
    \frac{T_{\rm a}}{S} = \frac{A_{\rm e}}{2k}
\end{equation}
where $k$ is the Boltzmann constant. Antenna gain expressed in this form is often referred to as ``degrees per flux unit'' (DPFU). Here we define it thus: 
\begin{equation}
    \label{eqn:gain}
    G_{\rm DPFU} \equiv \frac{A_{\rm e}}{2k}
\end{equation}
to distinguish it from the dimensionless gain $G$ while retaining some consistency with usage elsewhere (particularly in the pulsar literature) where a plain $G$ is often used. This quantity is typically rendered in units of K\,Jy$^{-1}$, where the jansky (Jy) 
corresponds to $10^{-26}$\,W\,m$^{-2}$\,Hz$^{-1}$. 
For the 64-m Parkes telescope $A_{p} = 3217$\,m$^2$ and so we can write:
\begin{equation}
\label{eqn:eap}
 \epsilon_{\rm ap} = 0.858[\rm Jy\,K^{-1}]\frac{T_{\rm sys}}{S_{\rm sys}}.
\end{equation}

Various other parameters relating to the antenna beam pattern can be defined. We define the main beam of the antenna as being out to the first zero and the main beam efficiency as:
\begin{equation}\label{eqn:mainBeam}
 \epsilon_B \equiv \frac{\Omega_{MB}}{\Omega_A}    
\end{equation}
where the main beam solid angle, $\Omega_{\rm MB}$, and the beam solid angle, $\Omega_A$:
\begin{equation}
    \Omega_{\rm MB, A} \equiv \frac{1}{G_{0}}\int_{\rm MB, A} G(\theta,\phi) d\Omega
\end{equation}
(giving $\Omega_A = 4\pi/G_{\rm 0}$).  Here $G_{\rm 0}$ is the peak gain of the antenna. We have presented estimates of the main beam efficiency for the UWL system as a function of frequency in Table~\ref{tb:subbands}.

The wide-band nature of the receiver system makes it difficult to estimate the signal-to-noise (S/N) of a pulsar observation because the pulsar's flux density, its pulse width, the system temperature and telescope gain all vary with frequency. The narrow-band spectral line sensitivity can be estimated using:
\begin{equation}
    \sigma_s = \frac{T_{\rm sys}}{G_{\rm DPFU}\sqrt{n_p \tau \Delta\nu}}
\end{equation}
where $\tau$ is the total on-source integration time, $\Delta\nu$ is the observing bandwidth and $n_p$ is the number of recorded polarisations (1 or 2).

\section{Data collection}

A publicly downloadable data collection is available from \url{https://doi.org/10.25919/5ce76bf409bcf}.  This data collection contains a ``README" file describing the contents of the data collection and directories that contain data sets relating to beam shapes (measured and simulated), antenna and beam efficiencies, the temperature determinations presented in Figure~\ref{fg:tsys} and the system equivalent flux densities shown in Figure~\ref{fg:tsys_ssys}.  We also provide flux and polarisation calibration data files and the data files corresponding to the astronomy examples (as shown in Figures~\ref{fg:spectrum} and \ref{fg:pulsar}).

\section{Author contributions}

The author list has an initial non-alphabetical section giving the names of people who have led specific aspects of the project. 

The remaining author list is in alphabetical order and includes the authors involved in building, installing and commissioning the receiver system.  EC, TE, VM, SO, TR and JS were part of the commissioning team.  BI provided input on the RFI sections of the paper.  MB, RB, JH, MK, AM, ED, NW, LW and WvS wrote the original funding grants.  WvS updated the MEM software and contributed to polarisation and flux calibration data analysis.

\end{appendix}

\bibliographystyle{pasa-mnras}
\bibliography{refs}

\end{document}